\documentclass{article}
\usepackage{amsmath,amssymb}

\newcommand{\hl}{H(\lambda)}
\newcommand{\hlt}{H(\lambda; t)}
\newcommand{\ho}{H_0}
\newcommand{\hi}{H_{\diamond}}
\newcommand{\uo}{U_0}
\newcommand{\uot}{U_0(t)}
\newcommand{\hm}{H_m}
\newcommand{\hn}{H_n}
\newcommand{\hil}{\hi (\lambda)}
\newcommand{\hilt}{\hi (\lambda ;t)}
\newcommand{\lt}{(\lambda ;t)}
\newcommand{\rotev}{T}
\newcommand{\zl}{Z(\lambda)}
\newcommand{\zlt}{Z(\lambda ;t)}
\newcommand{\lto}{(\lambda;t,t_0)}
\newcommand{\clt}{C(\lambda ;t)}
\newcommand{\cl}{C(\lambda)}
\newcommand{\ttt}{\mathfrak{t}}
\newcommand{\cltint}{\!\int_0^t\! C(\lambda ;\ttt)\, \de\ttt}
\newcommand{\rotham}{\tilde{H}}
\newcommand{\rothamlt}{\tilde{H}(\lambda ;t)}
\newcommand{\spa}{\!\!\!\!}
\newcommand{\de}{\mathrm{d}}
\newcommand{\der}{\frac{\mathrm{d}\ }{\mathrm{d}t}}
\newcommand{\Ad}{\mathrm{Ad}}
\newcommand{\ad}{\mathrm{ad}}
\newcommand{\X}{\mathfrak{X}}
\newcommand{\zz}{\mathcal{Z}}
\newcommand{\zad}{\breve{\mathcal{Z}}}
\newcommand{\rrc}{\lfloor C\rfloor}
\newcommand{\rrcad}{\lfloor\breve{C}\rfloor}
\newcommand{\cad}{\breve{C}}
\newcommand{\cadint}{\!\int_0^t\! \breve{C}(\lambda ;\ttt)\, \de\ttt}
\newcommand{\cc}{\mathfrak{C}}
\newcommand{\ccad}{\breve{\mathfrak{C}}}
\newcommand{\hb}{\overline{H}(\lambda)}
\newcommand{\JJJ}{\mathsf{J}}
\newcommand{\avxp}{\mathrm{avxp}}
\newcommand{\dexp}{\mathsf{d}\hspace{0.4mm}\mathrm{exp}}

\newcommand{\hG}{\hat{\mathsf{G}}}

\newcommand{\tauc}{{^{\tau}\!C}}
\newcommand{\tauz}{{^{\tau}\!Z}}
\newcommand{\enne}{\mathsf{N}}
\newcommand{\enneind}{_{\mbox{\tiny $[\mathsf{N}]$}}}
\newcommand{\app}{\stackrel{\ \lambda^\enne}{\approx}}
\newcommand{\infc}{{^{\infty}\hspace{-0.4mm}C}}
\newcommand{\infz}{{^{\infty}\!Z}}
\newcommand{\minc}{{^{\bowtie}\hspace{-0.2mm}C}}
\newcommand{\minz}{{^{\bowtie}\!Z}}

\newcommand{\extra}[1]{\left\vert\rangle#1\langle\right\vert_{\ho}}
\newcommand{\infra}[1]{\left\langle\vert#1\vert\right\rangle_{\!\ho}}
\newcommand{\extr}[1]{\left\vert\rangle#1\langle\right\vert}
\newcommand{\infr}[1]{\left\langle\vert#1\vert\right\rangle}
\newcommand{\Extra}[1]{\left[\vert\rangle#1\langle\vert\right]_{\ho}}
\newcommand{\extrat}[1]{\left\vert\left\rangle#1\right\langle\right\vert_{\ho(t)}}
\newcommand{\infrat}[1]{\left\langle\left\vert#1\right\vert\right\rangle_{\!\ho(t)}}
\newcommand{\infratt}[1]{\langle\vert#1\vert\rangle_{\!\ho(t)}}
\newcommand{\Extrat}[1]{\left[\left\vert\left\rangle#1\right\langle\right\vert\right]_{\ho(t)}}
\newcommand{\GG}{\mathcal{G}}
\newcommand{\GGG}{\mathsf{G}}
\newcommand{\BB}{\mathcal{B}}
\newcommand{\BBB}{\mathsf{B}}
\newcommand{\RR}{\mathcal{R}}
\newcommand{\RRR}{\mathsf{R}}
\newcommand{\TC}{T_C}
\newcommand{\TZ}{T_Z}
\newcommand{\zzz}{\mathfrak{Z}}
\newcommand{\period}{\mathtt{T}}
\newcommand{\kkato}{\mathcal{K}}
\newcommand{\katof}{K\hspace{-2.5mm}\overset{\smallfrown}{\phantom{a}}}
\newcommand{\katos}{K\hspace{-2.5mm}\overset{\smallsmile}{\phantom{a}}}
\newcommand{\kkatof}{\mathcal{K}\hspace{-2.3mm}\overset{\smallfrown}{\phantom{a}}}
\newcommand{\kkatos}{\mathcal{K}\hspace{-2.3mm}\overset{\smallsmile}{\phantom{a}}}
\newcommand{\adham}{\mathcal{J}\hspace{-3mm}\sim\hspace{-2.3mm}\mathcal{C}}
\newcommand{\tscale}{\mbox{\small $\mathcal{T}$}}

\begin{document}

\title{A perturbative expansion of the evolution operator
associated with time-dependent quantum Hamiltonians
\vspace{0.5cm}}

\author{
{\Large P. Aniello\,}\footnote{\; Paolo.Aniello@na.infn.it} \vspace{0.3cm}\\
{\small Istituto Nazionale di Fisica Nucleare, Sezione di
Napoli,}\\{\small and}\\
{\small Dipartimento di Scienze Fisiche dell'Universit\`{a} di
Napoli \lq\lq Federico II\rq\rq,}
\\
{\small Complesso Universitario di Monte S. Angelo, Via Cintia -
80126 Napoli, Italy} \vspace{0.4cm}}


\maketitle

\begin{abstract}
A novel expansion --- which generalizes Magnus expansion
--- of the evolution operator associated with a (in general, time-dependent)
perturbed Hamiltonian is introduced. It is shown that it has a wide range of
possible solutions that can be fitted according to computational
convenience. The time-independent and the adiabatic case are
studied in detail.
\end{abstract}

\vspace{3 cm}

\noindent Keywords: {perturbation theory}\\
PACS: 31.15.Md

\clearpage


\section{Introduction}

The explicit determination of the evolution operator associated with a
quantum system is a `touchy business'.
If the Hamiltonian of the system does not depend on time and has the form
of the sum of a solvable unperturbed Hamiltonian plus an analytic perturbation,
one can use the tools of standard perturbation
theory~{\cite{Kato-pert}\cite{Kato}} for linear operators,
based on the expansion of the resolvent, in order to achieve approximate
expressions of the evolutor, or apply a suitable unitary operator
perturbative approach~\cite{Aniello1}~\cite{Aniello2}~\cite{Militello}.\\
If the Hamiltonian is time-dependent (i.e.\ it describes a
non-isolated quantum system), the problem is in general even more
radical. In fact, it is well known that, whenever the values of
the Hamiltonian at different times do not commute, the evolutor
does not admit a simple formal expression.

In two fundamental papers~\cite{Dyson} in the history of quantum
electrodynamics, Dyson developed an expansion of the evolution
operator that has been adopted extensively in any field of
physics. Dyson expansion has a transparent physical interpretation
in terms of time ordered elementary processes which makes its
application particularly appealing, especially in quantum field
theory. On the other hand, for many applications, Dyson expansion
has severe drawbacks, as a low convergence rate and the lack of
unitarity of its
truncations~\cite{Pechukas}.\\
Later, Magnus~\cite{Magnus} introduced an expansion of the
evolution operator such that each of its truncations retains the
property of being unitary. Magnus expansion has been
`rediscovered' and re-elaborated several times (see for instance
ref.~\cite{Pechukas}), applied successfully to several problems
--- nuclear magnetic resonance~\cite{Evans}, atomic collision
theory~\cite{Baye}, molecular systems in intense laser
fields~\cite{Milfeld}, neutrino oscillations in
matter~\cite{D'Olivo}, to quote just a small sample --- and its
convergence properties have been
studied~\cite{Fernandez}~\cite{Moan}.

On our opinion, Magnus expansion, rather than Dyson expansion,
should be regarded as the most natural generalization of the
expression of the evolutor associated with a time-independent
Hamiltonian. In fact, it is written in the form of the exponential
of the expansion of a suitable time-dependent anti-hermitian
operator which can be deduced, order by order, from the
Hamiltonian of the system. Now, precisely for this reason, just
like for the evolutor generated by a time-independent Hamiltonian,
the problem of {\it computing} explicitly the action of (any
truncation of) the Magnus expansion on the state vectors is
non-trivial. Expanding the exponential would lead to non-unitary
truncations, thus to the loss of the most important feature of
Magnus expansion. Then, the issue of finding a generalization of
Magnus expansion retaining the property of having unitary
truncations, but allowing simpler explicit solutions, arises in a
natural way.

In the present paper, we have tried to achieve this result. Our
basic idea is simple: to combine the Magnus expansion with the
passage to a suitable interaction picture. Precisely, given a
perturbed Hamiltonian, after the usual passage to the interaction
picture that decouples the unperturbed dynamics (which is supposed
to be known explicitly), one switches to a further interaction
picture, depending on the perturbative parameter, in order to
achieve computational advantages. We stress that the idea of
`adapting' Magnus expansion is not new. It appears in a paper by
Casas~{\it et al.}~\cite{Casas} in which the authors introduce the
{\it Floquet-Magnus expansion} for the evolution operator
associated with a (interaction picture) Hamiltonian depending
periodically on time. Our approach generalizes the one proposed by
Casas~{\it et al.}\ even in the case when the
interaction picture Hamiltonian is periodic on time.\\
We have made the choice of skipping mathematical complications.
For instance, it is known that even a simple passage to an
interaction picture can be mathematically tricky (see, for
example, ref.~\cite{Simon}). Our choice is motivated by various
reasons. First of all, we believe that heuristic investigation
should always precede rigorous re-elaboration. Once that it is
clear what the basic `rules of the game' are, one can adopt the
most appropriate mathematical tools. Moreover, we avoid the risk
of hiding in a cloud of technicalities the main ideas and of
discouraging those physicists who may want to {\it apply} our method
for solving problems. It should be also observed that a recent
trend in quantum mechanics is to focus on systems which can be
described by effective Hamiltonians in finite-dimensional Hilbert
spaces (consider, in particular, the huge research area related to
quantum computation and quantum information theory; see
ref.~\cite{Galindo} and the rich bibliography therein). The study
of these systems is not affected by all the technicalities
associated with the infinite-dimensional spaces but retains all
the most intriguing features of quantum physics.

The paper is organized as follows. In section~{\ref{basic}}, we
introduce the basic decomposition of the evolution operator which
will allow us to obtain a perturbative expansion. Two important
cases --- the time-independent case and the adiabatic case ---
will be considered in sections~{\ref{independent}}
and~{\ref{solution}}. In section~{\ref{generalcase}}, we will
study the general time-dependent case and show how the solutions
obtained correspond, in the time-independent case, to the ones
obtained in section~{\ref{independent}}.

\section{Basic assumptions and strategy}
\label{basic}

Let us consider a time-dependent perturbed Hamiltonian $\hlt$,
namely a selfadjoint linear operator of the
form
\begin{equation} \label{hamiltonian}
\hlt=\ho(t)+\hilt,
\end{equation}
where $\ho(t)$ is a selfadjoint (and, in general, time-dependent)
operator --- the `unperturbed component' --- and $\hilt$ is a
time-dependent perturbation; precisely, we will assume that
$\lambda\mapsto\hilt$ is (for the perturbative parameter $\lambda$ in a certain
neighborhood of zero and for any $t$) a real analytic,
selfajoint, bounded
operator-valued function, with $\hi(0;t)=0$.
A real analytic function can be extended to a domain in the complex
plane. Keeping this fact in mind, we will specify that a given property
{\it holds for $\lambda$ real}. For instance, the analytic function $\lambda\mapsto\hilt$
will take values in the selfadjoint operators for $\lambda$ real only.

Let $U(\lambda;t,t_0)$ be the evolution operator associated with
$\hlt$, with initial time $t_0$; namely ($\hbar=1$):
\begin{equation}
i\,\dot{U}(\lambda;t,t_0)=\hlt\,U(\lambda;t,t_0),
\ \ \ U(\lambda ;t,t_0)=\mathrm{Id},
\end{equation}
where the dot denotes the time derivative. Then, we have that
\begin{equation}
U(\lambda;t,t_0)= \uo(t,t_0)\, \rotev(\lambda;t,t_0),
\end{equation}
where $\uo(t,t_0)$ and
$\rotev(\lambda;t,t_0)$
are respectively the evolution operator associated with the unperturbed
component $\ho(t)$ (evolution operator which, if the unperturbed Hamiltonian
is time-independent, $\ho(t)\equiv\ho$, is obviously given by
$e^{-i\ho (t-t_0)}$)
and the evolution operator associated with the
interaction picture Hamiltonian
\begin{equation}
\rotham(\lambda;t,t_0) := \uo(t_0,t)\, \hilt \, \uo(t,t_0).
\end{equation}
Let us notice explicitly that, since $\rotham(0;t,t_0)=0$, we have:
\begin{equation} \label{condi}
T(0;t,t_0)=\mathrm{Id}.
\end{equation}

We will suppose that the unperturbed evolution $\uo(t,t_0)$ is
explicitly known. Then the problem is to determine perturbative expressions
of $\rotev(\lambda;t,t_0)$. To this aim, the central point of the paper
is the assumption that $\rotev(\lambda;t,t_0)$ has the following
general form:
\begin{equation} \label{gen1}
\!\!\rotev(\lambda;t,t_0) = \exp\left(-i\,Z\lto\right)\,
\exp\!\left(\!\!-i\!\int_{t_0}^t\!C(\lambda;\ttt,t_0)\,\de\ttt\!\right)
\exp\left(i\,Z(\lambda;t_0,t_0)\right),
\end{equation}
where $(\lambda ;t)\mapsto Z\lto$, $(\lambda ;t)\mapsto C\lto$ are
operator-valued functions which depend analytically on the
perturbative parameter $\lambda$;
in agreement with condition~{(\ref{condi})}, we set:
\begin{equation}
Z(0;t,t_0)=0,\ \ \ C(0;t,t_0)=0,\ \ \ \ \forall t.
\end{equation}
We stress that the presence of
the term $\exp(i\,Z(\lambda; t_0,t_0))$ in formula~{(\ref{gen1})}
ensures that $T(\lambda;t_0,t_0)=\mathrm{Id}$,
allowing the possibility that $Z(\lambda;t_0,t_0)\neq 0$.

It will be seen that decomposition~{(\ref{gen1})} has a wide range
of solutions and that a possible choice for fixing a certain class
of solutions is given by imposing the condition $C\lto=\cl$, i.e.
assuming that the function $(\lambda ;t)\mapsto C\lto$ does not
depend on time. This decomposition includes, as particular cases,
two decompositions of the evolution operator that have been
considered in the literature:
\begin{itemize}
\item the decomposition that is obtained setting
\[
Z\lto=0,\ \ \ \ \forall t,
\]
in formula~{(\ref{gen1})}, decomposition
which is at the root of the {\it Magnus expansion} of the
evolution operator~{\cite{Magnus}};
\item the classical {\it Floquet decomposition} that holds in the
case where the interaction picture Hamiltonian depends periodically
on time (let us denote the period by $\period$) --- decomposition which
is obtained setting
\[
C\lto\equiv\cl, \ \ \ Z(\lambda;t_0,t_0)=0,
\]
and assuming that $(\lambda,t)\mapsto Z(\lambda;t,t_0)$ is
periodic with respect to time with period $\period$ --- and that is at
the root of the {\it Floquet-Magnus} expansion of the evolution
operator~{\cite{Casas}}.
\end{itemize}

From this point onwards, for notational convenience, we will fix
$t_0=0$. Then, decomposition~{(\ref{gen1})} can be rewritten as
\begin{equation} \label{gen2}
\rotev\lt =
\exp\left(-i\zlt\right)\,\exp\!\left(-i\cltint\right)\,\exp\left(i\zl\right),
\end{equation}
where we have set:
\begin{equation*}
\rotev\lt\equiv \rotev(\lambda;t,0),\ \zlt\equiv Z(\lambda;t,0),\
\zl\equiv Z(\lambda; 0), \ \clt\equiv C(\lambda;t,0).
\end{equation*}
We are now ready to obtain a perturbative expansion of
$\rotev\lt$. In fact, if we require the interaction picture
evolution operator to satisfy the Schr\"odinger equation, we get:
\begin{eqnarray}
\rothamlt\,\rotev\lt
\spa & = & \spa i\,\dot\rotev\lt
\nonumber \\
& = & \spa
e^{-i\zlt}\!\!\int_{0}^{1}\!\!\!\left(e^{is\zlt}\,\dot{Z}\lt\,e^{-is\zlt}\right)\!
\de s\ e^{-i\cltint} e^{i\zl} + \nonumber \\
& & \spa\spa\spa\spa\spa\spa\spa\spa\spa\spa +\
e^{-i\zlt}\!\!\int_{0}^{1}\!\!\!\left(e^{-is\cltint}\,\clt\,
e^{is\cltint}\right)\!
\de s\ e^{-i\cltint} e^{i\zl},
\label{preceding}
\end{eqnarray}
where we have used the remarkable formula (see, for instance,
ref.~\cite{Wilcox})
\begin{equation}
\der\, e^F = e^F \int_{0}^{1} \!\left(e^{-sF}\, \dot{F} \,
e^{sF}\right)\! \de s = \int_{0}^{1}\! \left(e^{sF}\, \dot{F}\,
e^{-sF}\right)\!\de s\  e^F ,
\end{equation}
which extends to an operator-valued function $t\mapsto F(t)$ the
standard formula for the derivative of the exponential of an
ordinary function. Next, let us apply to each member of
equation~{(\ref{preceding})} the operator $e^{i\zlt}$ on the left
and the operator $e^{-i\zl}e^{i\cl t}$ on the right; we find:
\begin{eqnarray}
\Ad_{\exp(i\zlt)}\,\rothamlt \spa & = & \spa \int_0^1 \!
\left(\Ad_{\exp(is\zlt)}\,\dot{Z}\lt\right)\!\de s
\nonumber\\
\label{prestart}
& + & \spa \int_0^1 \!
\left(\Ad_{\exp\left(-is\cltint\right)}\,\clt\right)\!\de s ,
\end{eqnarray}
where we recall that, given linear operators $\X,Y$, with $\X$
invertible,
\begin{equation}
\Ad_{\X}\,Y:=\X\, Y\,\X^{-1}.
\end{equation}
Then, since $\X$ is of the form $e^X$, we can use the well known
relation
\begin{equation} \label{form}
\Ad_{\exp(X)}\, Y=\exp(\ad_X)\, Y =\sum_{k=0}^\infty
\frac{1}{k!}\,\ad_X^k\, Y,
\end{equation}
with $\ad_X^k$ denoting the $k$-th power ($\ad_X^0\equiv\mathrm{Id}$)
of the superoperator $\ad_X$ defined by
\begin{equation}
\ad_X\,Y:=[X,Y].
\end{equation}
Eventually, applying formula~{(\ref{form})} to equation~{(\ref{prestart})}
and performing the integrations, we obtain:
\begin{eqnarray}
\sum_{k=0}^\infty \frac{i^k}{k!}\,\ad_{\zlt}^k\, \rothamlt \spa &
= & \spa \sum_{k=0}^\infty \frac{i^k}{(k+1)!}\,\ad_{\zlt}^k\,
\dot{Z}\lt \nonumber\\ \label{eqcompl} & + & \spa
\sum_{k=0}^\infty \frac{(-i)^k}{(k+1)!}\,\ad_{\cltint}^k\,\clt.
\end{eqnarray}
This equation will be the starting point for the determination of
the operator-valued functions $(\lambda,t)\mapsto\zlt$ and $(\lambda,t)
\mapsto\clt$ at each perturbative order in $\lambda$,
task that will be pursued sistematically in the next sections.


\section{The time-independent case and the perturbative
adiabatic approximation}
\label{independent}

In this section, we will consider two important cases:
\begin{enumerate}
\item the case where the Hamiltonian~{(\ref{hamiltonian})} does not depend
on time;
\item the case where the Hamiltonian~{(\ref{hamiltonian})} is slowly varying
with respect to time (so that the adiabatic approximation can be applied).
\end{enumerate}
As we will show later on, the second case, within the adiabatic
approximation, can be treated by a method analogous to the one
adopted in the first case. In both cases it will be convenient to
set
\begin{equation}
\zz\lt:=\uot\,\zlt\,\uot^\dagger,\ \ \ \zz(\lambda)\equiv\zz(\lambda;0)=\zl,
\end{equation}
and re-express equation~{(\ref{eqcompl})} in terms of the transformed
operator $\zz\lt$. To this aim, let us first notice that
\begin{equation} \label{derivative}
\dot{Z}\lt=\Ad_{\uot^\dagger}\left(\dot{\zz}\lt-i\,\ad_{\zz\lt}\,\ho\right).
\end{equation}
Besides, given linear operators $\X$, $X$ and $Y$, with $\X$ invertible,
one can show inductively that
\begin{equation} \label{relad}
\ad_{\Ad_{\X} X}^k\,\Ad_{\X}\, Y=\Ad_{\X}\,\ad_X^k\, Y,\ \ \
k=0,1,2,\ldots\ .
\end{equation}
Then, since $\zlt=\Ad_{\uot^\dagger}\,\zz\lt$,
$\rothamlt=\Ad_{\uot^\dagger}\,\hilt$
and relation~{(\ref{derivative})} holds, using formula~{(\ref{relad})},
from equation~{(\ref{eqcompl})} we obtain:
\begin{eqnarray}
\!\!\! \Ad_{\uot^\dagger}\sum_{k=0}^\infty
\frac{i^k}{k!}\,\ad_{\zz\lt}^k\,\hilt \spa & = & \spa
\Ad_{\uot^\dagger}\sum_{k=0}^\infty
\frac{i^k}{(k+1)!}\,\ad_{\zz\lt}^k\,\dot{\zz}\lt \nonumber\\
& - & \spa \Ad_{\uot^\dagger}\sum_{k=1}^\infty
\frac{i^k}{k!}\,\ad_{\zz\lt}^k\,\ho
\nonumber\\
& + & \spa  \sum_{k=0}^\infty
\frac{(-i)^k}{(k+1)!}\,\ad_{\cltint}^k\,\clt.
\end{eqnarray}
Next, applying the superoperator $\Ad_{\uot}$ to each
member of this equation and rearranging the terms, we get
\begin{eqnarray}
\spa  \sum_{k=1}^\infty
\frac{i^k}{k!}\,\ad_{\zz\lt}^k\!\left(\ho(t)+\hilt\right) +\hilt \spa
& = &
\spa \Ad_{\uot}\Big(\clt \nonumber\\
& + & \spa  \sum_{k=1}^\infty
\frac{(-i)^k}{(k+1)!}\,\ad_{\cltint}^k\,\clt\Big) \nonumber\\ & +
& \spa \sum_{k=0}^\infty
\frac{i^k}{(k+1)!}\,\ad_{\zz\lt}^k\,\dot{\zz}\lt. \label{gen3}
\end{eqnarray}

\subsection{The time-independent case}

In the {\it time-independent case}, we have that
$\ho(t)\equiv\ho$, $\hilt\equiv\hil$, and it is natural to set:
\begin{equation} \label{tindcond}
\clt=\cl,\ \ \ \zz\lt =\zz (\lambda ;0)=Z(\lambda ;0)\equiv\zl.
\end{equation}
Then, equation~{(\ref{gen3})} assumes a much simpler form:
\begin{equation} \label{cost}
\sum_{k=1}^\infty
\frac{i^k}{k!}\,\ad_{\zl}^n\!\left(\ho+\hil\right)+\hil = e^{-i\ho
t}\,\cl\, e^{i\ho t}.
\end{equation}
Now, observe that the first member of this equation does not
depend on time, hence the function $t\mapsto e^{-i\ho t}\,\cl\,
e^{i\ho t}$ must be constant. It follows that, if we want
equation~{(\ref{cost})} to be consistent, we have to assume that
\begin{equation} \label{cond}
[\cl,\ho]=0,
\end{equation}
i.e. that $\cl$ {\it is a constant of the motion for the
unperturbed evolution generated by} $\ho$.
Eventually, we obtain the following fundamental formula:
\begin{equation} \label{decomposition}
\sum_{k=1}^\infty
\frac{i^k}{k!}\,\ad_{\zl}^k\!\left(\ho+\hil\right) +\hil =
C(\lambda).
\end{equation}
At this point, we are ready to obtain perturbative expansions of
the operators $\cl$ and $\zl$ (hence, of the interaction picture
evolution operator $\rotev(\lambda ;t)$). We will assume that the
unperturbed Hamiltonian $\ho$ has a pure point spectrum, while the
case where this hypothesis is not satisfied is a particular case of
the general treatment developed in section~{\ref{generalcase}}
(indeed, in this case the formulae obtained in this section, which
involve eigenvalues and eigenprojectors, make no sense). We will
denote by $E_1,E_2,\ldots$ the (possibly degenerate) eigenvalues
of $\ho$ and by $P_1,P_2,\ldots$ the associated eigenprojectors.
Since the functions $\lambda\mapsto\hil$, $\lambda\mapsto
C(\lambda)$ and $\lambda\mapsto Z(\lambda)$ are analytic and
$\hi(0)=C(0)=Z(0)=0$, we can write:
\begin{equation} \label{power}
\hil = \sum_{n=1}^\infty \lambda^n\, H_n,\ \ \
C(\lambda)=\sum_{n=1}^{\infty}\lambda^n\, C_n,\ \ \
Z(\lambda)=\sum_{n=1}^{\infty}\lambda^n\, Z_n .
\end{equation}
Now, in order to determine the operators $\{C_n\}_{n\in\mathbb{N}}$ and
$\{Z_n\}_{n\in\mathbb{N}}$,
let us substitute the power expansions~{(\ref{power})} in
equation~{(\ref{decomposition})}; in correspondence to the various
orders in the perurbative parameter $\lambda$, we get the
following set of conditions:
\begin{eqnarray}
C_1 -i\left[Z_1,\ho\right]-H_1=0 \!\! & , & \!\!
\left[C_1,\ho\right]=0 \label{first}
\\
\spa\spa\spa
C_2-i\left[Z_2,\ho\right]+\frac{1}{2}\left[Z_1,\left[Z_1,\ho\right]\right]
-i\left[Z_1,H_1\right]-H_2=0 \!\! & , & \!\!
\left[C_2,\ho\right]=0 \label{second}
\\
& \vdots & \nonumber
\end{eqnarray}
where we have taken into account the additional constraint
$[C(\lambda),\ho]=0$. This infinite set of equations can be solved
recursively and the solution --- as it should be expected (we will
clarify this point soon) --- is not unique.
The first equation, together with the first constraint, determines
$Z_1$ up to an operator commuting with $\ho$ and $C_1$ uniquely.
Indeed, since
\begin{equation}
[C_1,\ho]=0\ \ \ \Rightarrow\ \ \
C_1=\sum_{m} P_m\,C_1\,P_m
\end{equation}
and
\begin{equation}
[Z_1,\ho]=\sum_{j\neq l} \left(E_l-E_j\right)P_j\,Z_1\,P_l,
\end{equation}
we conclude that
\begin{equation} \label{formulac}
C_1 = \sum_{m} P_m\,H_1\, P_m
\end{equation}
and
\begin{equation}
Z_1= \sum_{m} P_m\,Z_1\, P_m + i\sum_{j\neq l}
\left(E_l-E_j\right)^{-1} P_j\,H_1\,P_l.
\end{equation}
This last equation admits a {\it minimal solution} which is
obtained by imposing a further condition, namely
\[
P_m\, Z_1\, P_m =0 \ \ \ \ m=1,2,\ldots\ .
\]

For $n>1$, we will adopt an analogous reasoning. Indeed, given an
operator $X$, let us set
\begin{equation} \label{ef}
\GG_{n}(X; Z_1,\ldots ,Z_{n}) := \sum_{m=1}^{n}\, \frac{i^m}{m!}
\!\!\! \sum_{\ \ \ k_1+\cdots +k_m=n} \!\! \ad_{Z_{k_1}} \cdots\
\ad_{Z_{k_m}}\, X,
\end{equation}
with $n\ge 1$. Then we can define the operator function
\begin{eqnarray} \label{gi}
\GGG_n(\ho,\ldots ,\hn ;Z_1,\ldots ,Z_{n-1}) \spa & := & \spa
\sum_{m=0}^{n-1} \GG_{n-m}(\hm;Z_1,\ldots ,Z_{n-m})
\nonumber \\
& - & \spa i[Z_n,\ho]+ H_n \ \ \ \ \ \ n\ge 2.
\end{eqnarray}
At this point, one can show that the sequence of equations
generated by formula~{(\ref{decomposition})} is given by
\begin{eqnarray}
C_1 -i\left[Z_1,\ho\right]=H_1 \!\! & , & \!\!
\left[C_1,\ho\right]=0
\nonumber\\
& \vdots & \nonumber\\
\spa \spa \spa C_n - i\left[Z_n,\ho \right] = \GGG_n (\ho,\ldots
,\hn; Z_1,\ldots, Z_{n-1}) \!\! & , &  \!\! \left[C_n,\ho\right]=0
\ \ \ \ n\ge 2 \label{general}
\\
& \vdots & \nonumber
\end{eqnarray}
In order to write the general solution of this sequence of
equations, it will be convenient to introduce a shorthand
notation; given a linear operator $X$, we set:
\begin{eqnarray}
\infra{X} \spa & := & \spa \sum_m P_m\,
X\,P_m:=\infr{X}_{\!\{\!P_m\!\}},
\\
\extra{X} \spa & := & \spa X-\infra{X}=\sum_{j\neq l} P_j\, X\,
P_l:=\extr{X}_{\{\!P_m\!\}},
\\
\Extra{X} \spa & := & \spa
i \sum_{j\neq l}(E_l-E_j)^{-1} P_j\, X \, P_l.
\end{eqnarray}
Notice that for the superoperators $\infra{\cdot}$ and
$\extra{\cdot}$, which differently from the superoperator
$\Extra{\cdot}$ do not depend on the eigenvalues of $\ho$,  we
have introduced the respective alternative symbols
$\infr{\cdot}_{\!\{\!P_m\!\}}$ and $\extr{\cdot}_{\{\!P_m\!\}}$
that will be used in the following whenever a certain set of {\it
spectral projections} $\{P_m\}$ (in general, not {\it
eigenprojections})
of a selfadjoint operator are involved.\\
Now, assume that the first $n$ equations have been solved. Then,
the operator $\GGG_{n+1}(\ho,\ldots ,H_{n+1}; Z_1,\ldots,Z_n)$ is
known explicitly and hence
\begin{equation} \label{forc}
C_{n+1}= \infra{\GGG_{n+1}(\ho,\ldots ,H_{n+1};
Z_1,\ldots,Z_n)},
\end{equation}
\begin{equation} \label{forz}
\left[Z_{n+1},\ho\right]= i\extra{\GGG_{n+1}(\ho,\ldots
,H_{n+1}; Z_1,\ldots,Z_n)}.
\end{equation}
Again, this last equation determines $Z_{n+1}$ up to
an arbitrary operator $\infra{Z_{n+1}}$
commuting with $\ho$; in fact, we have:
\begin{equation}
Z_{n+1}=\infra{Z_{n+1}}+
\Extra{\GGG_{n+1}(\ho,\ldots,H_{n+1};Z_1,\ldots,Z_n)}.
\end{equation}
We stress that,
in general, the choice of a particular
solution for $Z_{n+1}$ will also
influence the form of $C_{n+2},Z_{n+2},\ldots\ $.

Thus, we conclude that the sequence of equations defined above
admits infinite solutions (even in the case where $\ho$ has a
nondegenerate spectrum). However, there is a unique {\it minimal
solution} $\{\minc_n,\minz_n\}_{n\in\mathbb{N}}$ which fulfills
the following additional condition:
\begin{equation} \label{minsol}
\infra{\minz_n}=0,\ \ \ \ n=1,2,\ldots\ .
\end{equation} \vspace{0.2cm}

In order to clarify the link of our approach with standard
perturbation theory for linear operators, let us recall a few
facts (see~{\cite{Kato}}~{\cite{Reed}}). It is possible to show
that, under certain technical conditions, there exist positive
constants $r_1,r_2,\ldots$ and a simply connected neighborhood
$\mathcal{I}$ of zero in $\mathbb{C}$ such that, for any
$\lambda\in\mathcal{I}$ and $m=1,2,\ldots\;$, one has that:
\vspace{0.4cm}

\begin{description}
\item[1)]\ \ \
the following
contour integral on the complex plane
\begin{equation} \label{proj}
P_m(\lambda)=\frac{1}{2\pi i}\oint_{\Gamma_m}\!\!\!\! dz \ \,
\left(z-\hl\right)^{-1}
\end{equation}
--- where $\Gamma_m$ is the anticlockwise oriented circle $[0,2\pi]\ni\theta\mapsto
E_m + r_m\, e^{i\theta}$ around the eigenvalue $E_m$
--- defines a projection ($P_m(\lambda)^2=P_m(\lambda)$), which is an
orthogonal projection for $\lambda\in\mathcal{I}\cap\mathbb{R}$,
with $P_m(0)=P_m$, and
$\mathcal{I}\ni\lambda\mapsto P_m(\lambda)$ is an analytic
operator-valued function;
\item[2)]\ \ \
the range of the projection $P_m(\lambda)$ is an invariant
subspace for $\hl$ (but, if the range of $P_m$ is not
1-dimensional, in general not an eigenspace), hence
\begin{equation} \label{inv}
\hl\, P_m(\lambda) = P_m(\lambda)\, \hl\, P_m(\lambda);
\end{equation}
\item[3)]\ \ \
there exists a (non-unique) analytic family $\lambda\mapsto
W(\lambda)$ of invertible operators such that
\begin{equation} \label{trasf}
P_m = W(\lambda)^{-1} P_m(\lambda) \, W(\lambda),\ \ \
W(0)=\mathrm{Id}
\end{equation}
--- with $W(\lambda)$ unitary for $\lambda$ real ---
which is solution of a Cauchy problem of the type
\begin{equation}
i\,W^\prime(\lambda)=J(\lambda)\,W(\lambda),\ \ \
W(0)=\mathrm{Id},
\end{equation}
where the apex denotes the derivative with respect to the
perturbative parameter and $\lambda\mapsto J(\lambda)$ is any
analytic family of operators --- selfadjoint for $\lambda$ real
--- such that
\begin{equation}
\extr{J(\lambda)}_{\{\!P_m(\lambda)\!\}} =i\sum_m
{P_m\hspace{-2.1mm}}^\prime\ (\lambda)\,P_m(\lambda)\ \ \ \ *\
P_m(\lambda)\, {P_m\hspace{-2.1mm}}^\prime\
(\lambda)\,P_m(\lambda)=0 \,*.
\end{equation}
\end{description}
In standard (Rayleigh-Schr\"odinger-Kato) perturbation theory, one can
obtain the perturbative corrections to unperturbed eigenvalues and
eigenvectors exploiting (see, for instance, ref.~{\cite{Messiah}})
formula~{(\ref{proj})} and the expansion of the resolvent operator
$(z-\hl)^{-1}$, namely
\begin{eqnarray}
\spa \spa (z-\ho-\hil)^{-1} \spa & = & \spa (z-\ho)^{-1}
-\lambda\,(z-\ho)^{-1} H_1\,(z-\ho)^{-1}
\nonumber\\
&-& \spa \lambda^2\,\Big( (z-\ho)^{-1} H_2\,(z-\ho)^{-1}
\nonumber\\
& + & \spa (z-\ho)^{-1}
H_1\,(z-\ho)^{-1}H_1\,(z-\ho)^{-1}\Big)+\ldots\ .
\end{eqnarray}
In our approach we use, instead, properties~{{\bf 2)} and {\bf
3)}. Indeed, let us define the operator $\hb$ by
\begin{equation}
\hb:=W(\lambda)^{-1} \hl\, W(\lambda),
\end{equation}
which, for real $\lambda$, is unitarily equivalent to $\hl$. Using
relations~{(\ref{inv})} and (\ref{trasf}), we find
\begin{eqnarray}
\hb\, P_m \spa & = & \spa W(\lambda)^{-1} \hl\, P_m(\lambda)\,
W(\lambda)
\nonumber\\
& = & \spa W(\lambda)^{-1} P_m(\lambda)\,\hl\,P_m(\lambda)\,
W(\lambda)
\end{eqnarray}
and hence:
\begin{equation}
\hb\, P_m= P_m \,\hb\, P_m \ \ \ \ m=1,2,\ldots\ .
\end{equation}
It follows that
\begin{equation}
\left[\,\hb,\ho\right]=0
\end{equation}
and then we obtain the following important relation:
\begin{equation}
\left[W(\lambda)^{-1} \hl\, W(\lambda) - \ho , \ho\right]=0.
\end{equation}
Thus, if we set
\begin{equation}
W(\lambda)^{-1} \hl\, W(\lambda) - \ho=\cl,\ \ \
W(\lambda)=\exp\!\left(-i\,\zl\right),
\end{equation}
and we apply relation~{(\ref{form})}, we find precisely
formula~{(\ref{decomposition})}.

Concluding our treatment of the time-independent case, it is worth
stressing that, due to conditions~{(\ref{tindcond})}, for the
overall evolution operator we have:
\begin{eqnarray}
U\lt \spa & = & \spa e^{-i\ho t}\, e^{-i\zlt}\, e^{-i\cl t}\,
e^{i\zl} \nonumber\\
*\, \zz\lt = e^{i\ho t}\,\zl\,e^{-i\ho t}\,* & = & \spa  e^{-i\zz\lt}\,
e^{-i\ho t}\, e^{-i\cl t}\, e^{i\zz(\lambda)}
\nonumber\\
*\, \left[\cl,\ho\right]=0\, * & = & \spa  e^{-i\zz\lt}\,
e^{-i\left(\ho +\cl\right)t}\, e^{i\zz(\lambda)},
\end{eqnarray}
or, more explicitly,
\begin{eqnarray}
U\lt \spa & = & \spa
e^{-i\zz\lt}\,\sum_m  e^{-i\left(E_m+\cl\right)t}\,P_m\, e^{i\zz(\lambda)}
\nonumber\\
& = & \spa
e^{-i\zz\lt}\,\sum_m \exp\!\left(-i(E_m+\rrc_m(\lambda))t\right) P_m\, e^{i\zz(\lambda)},
\end{eqnarray}
where we have introduced the `reduced rank operators'
\begin{equation}
\rrc_m(\lambda):=C(\lambda)\,P_m=P_m\,C(\lambda)\,P_m,\ \ \ m=1,2,\ldots\ .
\end{equation}

\subsection{The adiabatic approximation}

Let us now consider the case where the {\it perturbative adiabatic
approximation} can be applied. This approximation consists
essentially in partially neglecting the last term in the r.h.s.\
of equation~{(\ref{gen3})} --- the one involving the time derivative
of $(\lambda,t)\mapsto \zz\lt$ --- under suitable conditions.\\
Precisely, we
will assume that the unperturbed Hamiltonian has (instantaneously)
a pure point spectrum with time-independent eigenprojectors.
Namely, we will suppose that there exists a set of orthogonal
projectors $\{P_m\}_{m=1,2,\ldots}$ forming a resolution of the
identity --- $\mathrm{Id}=\sum_m P_m$ --- such that
it coincides with the set of eigenprojectors of $\ho(t)$, for any $t$.\\
This hypothesis prevents the possibility of occurence of `level
crossings' in the spectrum of the unperturbed Hamiltonian.
Indeed, it implies that there exist real functions
$t\mapsto E_1(t),\, t\mapsto E_2(t),\ldots$ such that, for any
$t$, $\{E_m(t)\}_{m=1,2,\ldots}$ is the set of the eigenvalues of
$\ho(t)$ --- specifically: $\ho(t)\, P_m=E_m(t)\, P_m$ --- and hence
\begin{equation}
E_1(t)\neq E_2(t)\neq\ldots\ \ \ \forall t .
\end{equation}
We will further assume that the functions $\{t\mapsto
E_m(t)\}_{m=1,2,\ldots}$ belong to
$\mathrm{C}^1(\mathbb{R})$.
Moreover, notice that --- since $[\ho(t),\ho(t^\prime)]=0,\ \forall
t,t^\prime$ ---
the unperturbed evolution operator will be given by:
\begin{equation} \label{commev}
\uot=e^{-i\int_0^t \ho(\ttt)\, \de\ttt}=\sum_m e^{-i\int_0^t
E_m(\ttt)\, \de \ttt}\, P_m.
\end{equation}
Then, we will consider a solution
$\{t\mapsto\cad\lt,\,t\mapsto\zad\lt\}$ of the equation
\begin{eqnarray}
\spa\spa\spa\sum_{k=0}^\infty
\frac{i^k}{k!}\,\ad_{\zad\lt}^k\!\left(\ho(t)+\hilt\right) \spa
&=& \spa \ho(t)+\cad\lt \nonumber\\ & + & \spa \sum_{k=1}^\infty
\frac{(-i)^k}{(k+1)!}\,\ad_{\cadint}^k\,\cad\lt \nonumber\\
\label{gen4} & + & \spa \infrat{\sum_{k=0}^\infty
\frac{i^k}{(k+1)!}\,\ad_{\zad\lt}^k\,\dot{\zad}\lt} ,
\end{eqnarray}
subject to the condition that
\begin{equation} \label{commute}
[\cad\lt,P_m]=0\ \ \ \ \forall t,\ \forall m.
\end{equation}
and
\begin{equation} \label{small}
\extrat{\sum_{k=0}^\infty
\frac{i^k}{(k+1)!}\,\ad_{\zad\lt}^k\,\dot{\zad}\lt}\!\!
\simeq 0.
\end{equation}
We observe explicitly that such a solution may not exist.
In more detail, one can show that a solution of equation~{(\ref{gen4})}
verifying condition~{(\ref{commute})} alone always exists (and, as in
the time-independent case, it is not unique),
but, in general,
not a solution verifying condition~{(\ref{small})} too; we do not insist
on this point here since it will be clarified afterwards.
We will then say that {\it the perturbative adiabatic approximation is
applicable} if a solution of equation~{(\ref{gen4})} verifying  both
conditions~{(\ref{commute})} and~{(\ref{small})} does exist.

Condition~{(\ref{commute})} is analogous to condition~{(\ref{cond})}
which has been assumed
in the time-independent case. Indeed, it is equivalent to the condition that
\begin{equation}
[\cad\lt,\ho(t^\prime)]=0\ \ \ \ \forall t,\ \forall t^\prime.
\end{equation}
Integrating with respect to time relation~{(\ref{commute})},
we find also that
\begin{equation} \label{commutint}
\left[\,\cadint\, ,P_m\right]\! =0\ \ \ \ \forall t,\ \forall m.
\end{equation}
Then, as a consequence of relation~{(\ref{commute})}, recalling
formula~{(\ref{commev})}, we have that
\begin{equation}
\left[\cad\lt,\uo(t^\prime)\right]\! =0,\ \ \ \left[\,\cadint\,
,\uo(t^\prime)\right]\!=0\ \ \ \ \forall t,t^\prime ;
\end{equation}
therefore:
\begin{equation} \label{commm}
\left[\uot , \cad\lt + \sum_{k=1}^\infty
\frac{(-i)^k}{(k+1)!}\,\ad_{\cadint}^k\,\cad\lt\!\right]\! = 0\ \
\ \ \forall t.
\end{equation}

Now, observe that if there exists a solution
$\{t\mapsto\cad\lt,\,t\mapsto\zad\lt\}$ of equation~{(\ref{gen4})}
verifying condition~{(\ref{commute})} (hence
relation~{(\ref{commm})}) and condition~{(\ref{small})}, then
equation~{(\ref{gen3})} will be approximately satisfied setting
$\clt=\cad\lt$ and $\zlt=\zad\lt$. Thus equation~{(\ref{gen4})}
will be the starting point for obtaining a perturbative expansion
of the evolution operator when the perturbative adiabatic
approximation is applicable. This task will be pursued in the next
section. Since at this stage the approximation considered may appear as
a mere {\it ad hoc} computational expedient, we will devote
the last part of this section to showing what is
its meaning; in particular, why we call it `adiabatic'.\\
We will first show that the evolution operator
$\breve{U}\lt$ associated with the
solution $\{t\mapsto\cad\lt,\, t\mapsto\zad\lt\}$ {\it behaves} like an
adiabatic evolutor.
To this aim, let us set
\begin{equation}
W\lt:=\exp\!\left(-i\,\zad\lt\right),
\end{equation}
and let us define the
projection
\begin{equation} \label{projlt}
P_m\lt := W\lt\, P_m \, W\lt^{-1}
\end{equation}
and the selfadjoint operators
\begin{equation} \label{hermitian}
\ccad\lt:= \cad\lt+\sum_{k=1}^\infty
\frac{(-i)^k}{(k+1)!}\,\ad_{\cadint}^k\,\cad\lt ,
\end{equation}
\begin{eqnarray} \label{Kkato}
\kkato\lt:=i\,W\lt^{-1}\dot{W}\lt \spa & = & \spa\sum_{k=0}^\infty
\frac{i^k}{(k+1)!}\,\ad_{\zad\lt}^k\,\dot{\zad}\lt
\nonumber\\
& = & \spa \kkatos\lt + \kkatof\lt,
\end{eqnarray}
with
\begin{equation}
\left\{
\begin{aligned}
\kkatos\lt := &\  \infrat{\sum_{k=0}^\infty
\frac{i^k}{(k+1)!}\,\ad_{\zad\lt}^k\,\dot{\zad}\lt} ,
\\
\kkatof\lt := &\ \extrat{\sum_{k=0}^\infty
\frac{i^k}{(k+1)!}\,\ad_{\zad\lt}^k\,\dot{\zad}\lt} .
\end{aligned}
\right.
\end{equation}
The operator $\ccad\lt$ (by virtue of relations~{(\ref{commute})}
and~{(\ref{commutint})}) and the operator $\kkatos\lt$ (by
definition) satisfy:
\begin{equation} \label{commute2}
[\ccad\lt, P_m]=0,\ \ \ [\kkatos\lt, P_m]=0,\ \ \ \ \forall t,\
\forall m.
\end{equation}
Then, equation~{(\ref{gen4})} can be rewritten as
\begin{equation} \label{decompad}
\hlt=W\lt\left(\ho(t)+\ccad\lt+\kkatos\lt \right) W\lt^{-1}
\end{equation}
and, by this equation, we have:
\begin{eqnarray}
\hlt\,P_m\lt \spa & = & \spa W\lt\left(
\ho(t)+\ccad\lt+\kkatos\lt\right) P_m
\,W\lt^{-1} \nonumber\\
*\ \mbox{relations (\ref{commute2})}\,* & = & \spa W\lt\, P_m
\left( \ho(t)+\ccad\lt+\kkatos\lt\right) P_m\,W\lt^{-1} \nonumber \\
& = & \spa P_m\lt\,\hlt\,P_m\lt,
\end{eqnarray}
i.e.\ {\it the range of the projection $P_m\lt$ is an invariant
subspace for} $\hlt$. At this point, it is convenient to make a detour.
\vspace{5mm}

We remark that as in the time-independent case --- instead of {\it
assuming} that decomposition~{(\ref{decompad})} holds --- one can {\it prove} that,
under suitable conditions, there exists a complete set $\{P_m\lt)\}_{m=1,2,\ldots}$
of spectral projections of $\hlt$ such that the functions
$\lambda\mapsto P_m\lt$ and $t\mapsto P_m\lt$ are
analytic and
\begin{equation}
P_m(0;0)=P_m,\ \ \ \ m=1,2,\ldots\ .
\end{equation}
Then, there is a (non-unique) unitary operator $W\lt$ such that
\begin{equation} \label{eqwlt}
P_m\lt=W\lt\,P_m\,W\lt^{-1},\ \ \ \ \forall t,\ \forall m.
\end{equation}
The operator $W\lt$ can be decomposed as
\begin{equation}
W\lt=A\lt\,W(\lambda),
\end{equation}
where $\lambda\mapsto W(\lambda)$ and $(\lambda, t)\mapsto A\lt$
are unitary operator-valued functions such that
\begin{eqnarray} \label{eqpm}
P_m(\lambda;0) \spa & = & \spa
W(\lambda)\,P_m\,W(\lambda)^{-1},
\\ \label{katoev}
P_m\lt \spa & = & \spa A\lt\,P_m(\lambda;0)\,A\lt,\ \ \ \ \forall
t,\ \forall m .
\end{eqnarray}
One can show that
these functions are solutions of Cauchy problems of the type
\begin{equation} \label{cauchy1}
i\,W^\prime(\lambda)=J(\lambda)\,W(\lambda),\ \ \ \ W(0)=\mathrm{Id},
\end{equation}
\begin{equation} \label{cauchy2}
i\,\dot{A}\lt=K\lt\,A\lt,\ \ \ \ A(\lambda;0)=\mathrm{Id} ,
\end{equation}
with $J(\lambda)$ and $K\lt$ selfajoint operators satisfying the
following conditions:
\begin{eqnarray}
\extr{J(\lambda)}_{\{\!P_m(\lambda;0)\!\}} \spa & = & \spa i\sum_m
{P_m\hspace{-2.1mm}}^\prime\ (\lambda;0)\, P_m(\lambda;0)
\nonumber\\ \label{condj} *\ 0=\frac{\mathrm{d}\
}{\mathrm{d}\lambda}\mathrm{Id}= \frac{\mathrm{d}\
}{\mathrm{d}\lambda}\sum_m P_m\lt^2\, * & = & \spa -i\sum_m
P_m(\lambda;0)\,{P_m\hspace{-2.1mm}}^\prime\ (\lambda;0),
\\
\katof\lt:=\extr{K\lt}_{\{\!P_m\lt\!\}} \spa & = & \spa
i\sum_{m}\dot{P}_m\lt\, P_m\lt \nonumber\\ \label{condk} *\
0=\der\mathrm{Id}=\der\sum_m P_m\lt^2\, * & = & \spa
-i\sum_{m}P_m\lt\,\dot{P}_m\lt .
\end{eqnarray}
Conversely, if $\lambda\mapsto W(\lambda)$ and $(\lambda, t)\mapsto A\lt$ are
operator-valued functions which are solutions of the Cauchy problems~{(\ref{cauchy1})}
and~{(\ref{cauchy2})} (with $J(\lambda)$ and $K\lt$ selfadjoint
operators subject to conditions~{(\ref{condj})} and~{(\ref{condk})}),
then they will satisfy equations~{(\ref{eqpm})} and~{(\ref{katoev})}.\\
In fact, suppose that relation~{(\ref{katoev})} is satisfied (we
will only argue for $A\lt$ since the argument for $W(\lambda)$ is
analogous). Then, setting $K\lt=i\,\dot{A}\lt\,A\lt^{-1}$, we
have:
\begin{equation} \label{eqk}
i\,\dot{P}_m\lt=[K\lt,P_m\lt],\ \ \ \forall t,\ \forall m.
\end{equation}
Conversely, if this relation holds, we have:
\begin{eqnarray}
\dot{\overbrace{A\lt^{-1} P_m\lt\,A\lt}} \spa & = & \spa A\lt^{-1}
\dot{P}_m\lt\,A\lt  + A\lt^{-1} P_m\lt\,\dot{A}\lt \nonumber\\
& - & \spa
A\lt^{-1}\dot{A}\lt\,A\lt^{-1} P_m\lt\,A\lt \nonumber \\
& = & \spa 0,\ \ \ \ \forall t,\ \forall m,
\end{eqnarray}
where we have used the fact that
$\dot{\overbrace{A\lt^{-1}}}=-A\lt^{-1}\dot{A}\lt$; hence:
\begin{equation}
\!\!\! A\lt^{-1} P_m\lt\,A\lt=A(\lambda;0)^{-1}
P_m(\lambda;0)\,A(\lambda;0)=P_m(\lambda;0), \ \, \forall t,\,
\forall m.
\end{equation}
Thus, relation~{(\ref{katoev})} holds if and only if $K\lt$
satisfies equation~{(\ref{eqk})}. At this point, using the fact
that $P_m\lt\,\dot{P}_m\lt\,P_m\lt=0,\ \forall m$, one can check
easily that
\begin{equation}
K\lt=\katos\lt+\katof\lt,
\end{equation}
--- with $\katos\lt :=\infr{K\lt}_{\hlt}$ and
$\katof\lt :=\extr{K\lt}_{\hlt}$ --- solves equation~{(\ref{eqk})}
if and only if:
\begin{equation}
\katof\lt=i\sum_m \dot{P}_m\lt\,P_m\lt.
\end{equation}
In conclusion, it is proven that, if there exists a complete set
of spectral projections $\{P_m\lt\}_{m=1,2,\ldots}$ of $\hlt$ with
the properties specified above, then equation~{(\ref{eqwlt})} is
satisfied by only and all the unitary operator-valued functions
$(\lambda, t)\mapsto W\lt$ such that $W(\lambda)=W(\lambda;0)$ and
$A\lt=W(\lambda)^{-1}W\lt$ are solutions of the Cauchy
problems~{(\ref{cauchy1})} and~{(\ref{cauchy2})}, with
conditions~{(\ref{condj})} and~{(\ref{condk})}. Moreover, if one
sets
\begin{equation}
\overline{H}\lt:=W\lt^{-1}\hlt\,W\lt,
\end{equation}
the following relation holds:
\begin{equation}
\overline{H}\lt\,P_m=P_m\,\overline{H}\lt\,P_m,\ \ \ \ \forall t,\
\forall m.
\end{equation}
Then, setting
\begin{eqnarray}
\kkatos\lt\spa & = & \spa i\infrat{W\lt^{-1}\dot{W}\lt}, \\
\ccad\lt\spa & = & \spa \overline{H}\lt-\ho(t)-\kkatos\lt,
\end{eqnarray}
one finds that
\begin{equation} \label{commcc}
[\ccad\lt,P_m]=0,\ \ \ \ \forall t,\ \forall m.
\end{equation}
In the next section, it will be shown that equation~{(\ref{hermitian})} allows to recover
the operator
$\cad\lt$ from $\ccad\lt$. Thus one can actually {\it define} $\cad\lt$
through formula~{(\ref{hermitian})}. One can also prove that if
relation~{(\ref{commcc})} holds then, as a {\it consequence}, $[\cad\lt,P_m]=0,\
\forall t,\,\forall m$. Hence, setting $W\lt=\exp(-i\,\zad\lt)$, one reobtains
equation~{(\ref{gen4})} with condition~{(\ref{commute})} automatically satisfied.
\vspace{5mm}

Let us now come back to our original purpose of investigating the behaviour
of the evolution operator $\breve{U}\lt$ that approximates $U\lt$:
\begin{eqnarray}
U\lt \spa & = & \spa \uot\, e^{-i\zlt}\, e^{-i\cltint}\, e^{i\zl}
\nonumber\\
\spa *\,  \zlt=\Ad_{\uot^\dagger}\,\zz\lt\, *\!\! & = & \spa
e^{-i\zz\lt}\,\uot \, e^{-i\cltint}\, e^{i\zz(\lambda)} \nonumber\\
*\  \mbox{adiabatic approximation} \, *\!\! & \simeq & \spa
e^{-i\zad\lt}\,\uot \, e^{-i\cadint}\, e^{i\zad(\lambda)}
\nonumber\\ & =: & \spa \breve{U}\lt .
\end{eqnarray}
Recalling formula~{(\ref{commev})} and
relation~{(\ref{commutint})}, the {\it adiabatic evolutor} $\breve{U}\lt$
can be written as
\begin{equation}
\breve{U}\lt= e^{-i\zad\lt}\,\sum_{m}
e^{-i\int_0^t\left(E_m(\ttt)+\cad(\lambda;\ttt)\right)\,\de\ttt}
\, P_m\, e^{i\zad(\lambda)},
\end{equation}
or --- introducing the reduced rank operator
\begin{equation}
\rrcad_m\lt:=\cad\lt\,P_m=P_m\,\cad\lt\,P_m,\ \ \ m=1,2,\ldots\
\end{equation}
--- in the more expressive form:
\begin{equation}
\breve{U}\lt= e^{-i\zad\lt}\,\sum_{m}
e^{-i\int_0^t\left(E_m(\ttt)+\rrcad_m(\lambda;\ttt)\right)\,\de\ttt}
\, P_m\, e^{i\zad(\lambda)}.
\end{equation}
To show that $\breve{U}\lt$ behaves indeed as an adiabatic
evolutor, let us observe that
\begin{eqnarray}
\breve{U}\lt\, P_j(\lambda;0) \spa & = & \spa
e^{-i\zad\lt}\,\uot\, e^{-i\cadint} \,  e^{i\zad(\lambda)}\,
P_j(\lambda;0)
\nonumber\\
\spa *\, P_j(\lambda;0)= \Ad_{e^{-i\zz(\lambda)}} P_j  \,*\!\! & =
& \spa e^{-i\zad\lt}\,\sum_m
e^{-i\int_0^t\left(E_m(\ttt)+\cad(\lambda;\ttt)\right)\,\de\ttt}\,P_m
\, P_j\, e^{i\zad(\lambda)}
\nonumber\\
& = & \spa e^{-i\zad\lt}\, P_j\,
e^{-i\int_0^t\left(E_j(\ttt)+\cad(\lambda;\ttt)\right)\,\de\ttt}
\, e^{i\zad(\lambda)}
\nonumber\\
& = & \spa e^{-i\zad\lt}\,P_j\,\uot\, e^{-i\cadint} \,
e^{i\zad(\lambda)}\nonumber \\
*\,\, \mbox{definition~{(\ref{projlt})}} \,
* \!\! & = & \spa P_j\lt\, \breve{U}\lt,\ \ \ j=1,2,\ldots\ ;
\end{eqnarray}
namely, $\breve{U}\lt$, as it should, intertwines the projection $P_j(\lambda;0)$
with the projection $P_j\lt$.
Let us do the following observations:
\begin{itemize}

\item The evolutor $\breve{U}\lt$ intertwines {\it spectral
projections}, that in general, for $\lambda\neq 0$, are not eigenprojections.
This situation is more general than the one considered originally by Kato~\cite{Kato-art} and
earlier by Born and Fock in their seminal paper~\cite{Born}.
Nevertheless, due to its importance in several applications, this situation has been
studied in later times by other authors (see, for instance, Nenciu's paper~\cite{Nenciu}).
Considering this more general situation is in our case unavoidable,
since in presence of the perturbation (i.e.\ for $\lambda\neq 0$)
the unperturbed eigenvalues (`energy levels') can `split'.

\item For $\lambda=0$, the adiabatic evolutor reduces to the
unperturbed evolution operator:
\begin{equation}
\breve{U}(0;t)=\uot.
\end{equation}
This fact justifies the term `{\it perturbative} adiabatic approximation'.

\end{itemize}

Anyway, in order to better highlight
the typical structure of an adiabatic evolutor, it is convenient
to rewrite the expression of $\breve{U}\lt$ as follows:
\begin{equation}
\breve{U}\lt=A\lt\,\sum_m
e^{-i\int_0^t(E_m(\ttt)+\Omega_m(\lambda;\ttt))\,\de\ttt}
\,P_m(\lambda;0),
\end{equation}
where
\begin{equation}
\left\{
\begin{aligned}
A\lt := &\  W\lt\,W(\lambda;0)^{-1},
\\
\Omega_m\lt := &\
W(\lambda;0)\,\breve{C}\lt\,W(\lambda;0)^{-1}P_m(\lambda;0)\\
= &\ W(\lambda;0)\,\rrcad_m\lt\,W(\lambda;0)^{-1}.
\end{aligned}
\right.
\end{equation}
Assume that the time-dependence of the Hamiltonian $\hlt$ is
characterized by a time scale $\tscale>0$, i.e.
\begin{equation}
\ho(t)=\adham_0(t/\tscale),\ \
\hilt=\adham_\diamond(\lambda;t/\tscale),\ \ \ \ t\in [0,\tscale],
\end{equation}
with ${\displaystyle [0,1]\ni s\mapsto \adham_0(s)}$ and
${\displaystyle [0,1]\ni s\mapsto \adham_\diamond(\lambda;s)}$
given operator-valued functions of the scaled time~$s$. From the
physicist's point of view, the parameter $\tscale$ measures the
`slowness' with which the non-isolated quantum system described by
the Hamiltonian $\hlt$ is influenced by the external world. Then,
for the spectral projections $\{P_m\lt\}_{m=1,2,\ldots}$ of the
Hamiltonian $\hlt$ we have:
\begin{equation}
P_m\lt=\mathcal{P}_m(\lambda;t/\tscale),\ \ \ \ \forall t\in
[0,\tscale],\ \forall m,
\end{equation}
where $\mathcal{P}_m(\lambda;s)$ is a projection-valued function of the
scaled time. \\
Recall now that
the operator $A\lt$ is solution of an initial value problem of the form
\begin{equation}
i\,\dot{A}\lt=K\lt\,A\lt,\ \ \ A(\lambda;0)=\mathrm{Id},
\end{equation}
where $K\lt$ has to satisfy the following condition:
\begin{eqnarray}
\extr{K\lt}_{\{\!P_m\lt\!\}} \spa & := & \spa i\sum_m \dot{P}_m\lt\,P_m\lt
\nonumber\\
*\; \dot{\mathcal{P}}_m(\lambda;s)\equiv \partial_s
\mathcal{P}_m(\lambda;s) \, *
& = & \spa
\frac{1}{\tscale}\, i
\sum_m \dot{\mathcal{P}}_m(\lambda;t/\tscale)\,\mathcal{P}_m(\lambda; t/\tscale).
\end{eqnarray}
Besides, as $K\lt=i\,\dot{W}\lt\,W\lt^{-1}$, recalling definition~{(\ref{Kkato})}
we find that the
selfadjoint operators $K\lt$ and $\kkato\lt$ are unitarily equivalent
since they are linked by the following formula:
\begin{equation}
\kkato\lt=W\lt^{-1} K\lt\,W\lt.
\end{equation}
More specifically, using the fact that
\begin{equation}
\! \extrat{\Ad_{W\lt^{-1}}(\cdot)}\!\!\equiv
\extr{\Ad_{W\lt^{-1}}(\cdot)}_{\{\!P_m\!\}}
\!\!=\Ad_{W\lt^{-1}}\!\extr{\cdot}_{\{\!P_m\lt\!\}},
\end{equation}
we have:
\begin{equation}
\kkatof\lt=W\lt^{-1}\katof\lt\,W\lt=
O\!\left(\frac{1}{\tscale}\right).
\end{equation}
This observation `completes the picture'. Indeed, it turns out
that, in the adiabatic limit, the contribution of the operator
$\kkatof\lt$ in equation~{(\ref{gen3})} can be neglected, namely
that approximation~{(\ref{small})} is justified as it is
equivalent to the standard adiabatic approximation.


\section{Solution with the perturbative adiabatic approximation}
\label{solution}

Let us now face the task of providing the perturbative solutions
of equation~{(\ref{gen4})}, which we recall is subject to
condition~{(\ref{commute})} and to what we can at this point
legitimately call `adiabatic approximation', i.e.\
condition~{(\ref{small})}.\\
It will be convenient to express
equation~{(\ref{gen4})} in terms of the operator $\ccad\lt$ in
place of $\cad\lt$. In fact, as it will be shown later on in this
section, one can develop a simple perturbative procedure which
allows to compute, order by order, the operator $\cad\lt$ from
$\ccad\lt$. Then, the equation that has to be solved
perturbatively is the following:
\begin{eqnarray}
\spa\spa\spa \sum_{k=1}^\infty
\frac{i^k}{k!}\,\ad_{\zad\lt}^k\!\left(\ho(t)+\hilt\right) \spa &
= & \spa \ccad\lt -\hilt \nonumber\\ \label{gen5} & + & \spa
\infrat{\sum_{k=0}^{\infty}\frac{i^k}{(k+1)!}\,\ad_{\zad\lt}^k\dot{\zad}\lt}\!\!,
\end{eqnarray} \vspace{-3mm}

\noindent where the operator $\cc\lt$ is subject to the costraint
\begin{equation} \label{costraint}
\left[\cc\lt,\ho(t)\right]=0, \ \ \ \ \forall t.
\end{equation}
It will be also shown that condition~{(\ref{costraint})} is
actually equivalent to condition~{(\ref{commute})} of which it is
a straightforward consequence. This fact is not immediately
evident from the definition of $\ccad\lt$. We will postpone the
problem of checking the validity of the adiabatic approximation
(condition~{(\ref{small})}) to later analysis.

Given linear operators $X,X_1,\ldots, X_{n}$ and
$Y_1,\ldots,Y_{n}$, let us set
\begin{equation}
\!\! \RR_{n}^\pm(X; Y_1,\ldots ,Y_{n}) := \sum_{m=1}^{n}\,
\frac{(\pm i)^m}{(m+1)!} \!\!\!\!\!\! \sum_{\ \ \ k_1+\cdots
+k_m=n} \!\!\!\!\!\!\! \ad_{Y_{k_1}} \cdots\ \ad_{Y_{k_m}}\, X, \;
\ \  n\ge 1.
\end{equation}
Then, for $n\ge 2$, we can define the operator function
\begin{equation} \label{defRRR}
\RRR_n^\pm(X_1,\ldots ,X_{n-1};Y_1,\ldots ,Y_{n-1}) :=\pm
\sum_{m=1}^{n-1} \RR_{n-m}^\pm(X_m;Y_1,\ldots ,Y_{n-m}).
\end{equation}
Noting the analogy of equation~{(\ref{gen5})} with
equation~{(\ref{decomposition})} obtained in the time-independent
case, we conclude that, given the power expansions
\begin{equation} \label{power2}
\!\! \hilt = \sum_{n=1}^\infty \lambda^n\, H_n(t),\ \,
\ccad\lt=\sum_{n=1}^{\infty}\lambda^n\, \ccad_n(t),\ \,
\zad\lt=\sum_{n=1}^{\infty}\lambda^n\, \zad_n(t) ,
\end{equation}
the sequence of equations generated by formula~{(\ref{gen5})},
with condition~{(\ref{costraint})}, has the following form:
\begin{eqnarray}
\ccad_1(t) - i\left[\zad_1(t),\ho(t) \right] \spa & = &  \spa
H_1(t)-\infrat{\dot{\zad}_1(t)},
\nonumber \\
\left[\ccad_1(t), \ho(t)\right] \spa & = & \spa 0,
\\
& \vdots & \nonumber
\\ \label{general*}
\ccad_n(t) - i\left[\zad_n(t),\ho(t) \right] \spa
& = &  \spa \GGG_n (\ho(t),\ldots ,\hn(t); \zad_1(t),\ldots,
\zad_{n-1}(t))
\nonumber \\
& - & \spa \infrat{\RRR_n^+(\dot{\zad}_1(t),\ldots
,\dot{\zad}_{n-1}(t);
\zad_1(t),\ldots ,\zad_{n-1}(t))} \nonumber\\
& - & \spa \infrat{\dot{\zad}_n(t)},
\nonumber \\
\left[\ccad_n(t),
\ho(t)\right] \spa & = & \spa 0 \ \ \ \ \ \ \ \ n\ge 2,
\\
& \vdots & \nonumber
\end{eqnarray}
Exactly as in the time-independent case, this system of equations
can be solved recursively. Indeed, we have that
\begin{eqnarray}
\ccad_{1}(t)\spa & = & \spa
\infrat{H_{1}(t)}\!-\der\infratt{\zad_1(t)},  \\
\zad_{1}(t) \spa & = & \spa
\infratt{\zad_1(t)}\!+ \Extrat{H_{1}(t)},
\end{eqnarray}
--- where $\infratt{\zad_1(t)}$ is an arbitrary operator commuting with
$\ho(t)$ (notice that ${\displaystyle \der\infratt{\zad_1(t)}=\infratt{\dot{\zad}_1(t)}}$) ---
and, assuming that the first $n$ equations have been
solved, so that the operator functions
\[
\GGG_{n+1}(\ldots ,H_{n+1}(t); \ldots,\zad_n(t)) \ \ \mbox{and}\ \
\RRR_{n+1}^+(\ldots,\dot{\zad}_{n}(t);\ldots,\zad_{n}(t))
\]
are known explicitly, the
solution of the $(n\!+\!1)$-th equation is given by
\begin{eqnarray}
\ccad_{n+1}(t) \spa & = & \spa \infrat{\GGG_{n+1}(\ho(t),\ldots ,H_{n+1}(t);
\zad_1(t),\ldots,\zad_n(t))} \nonumber\\
& - & \spa
\infrat{\RRR_{n+1}^+(\dot{\zad}_1(t),\ldots,\dot{\zad}_{n}(t);\zad_1(t),\ldots,\zad_{n}(t))}
\nonumber\\ \label{forc*} & - & \spa \der\infrat{\zad_{n+1}(t)},
\end{eqnarray}
\begin{equation} \label{forz*}
\!\! \left[\zad_{n+1}(t),\ho(t)\right]\!=
i\extrat{\GGG_{n+1}(\ho(t),\ldots ,H_{n+1}(t);
\zad_1(t),\ldots,\zad_n(t))}\!\!.
\end{equation}
Again, equations~{(\ref{forc*})} and~{(\ref{forz*})} determine
$\ccad_{n+1}(t)$ and $\zad_{n+1}(t)$ up to an
operator $\infratt{\zad_{n+1}(t)}$ commuting with $\ho(t)$ and we
have:
\begin{eqnarray}
\spa\zad_{n+1}(t)\spa & = & \spa \infrat{\zad_{n+1}(t)} \nonumber\\
& + & \spa
\Extrat{\GGG_{n+1}(\ho(t),\ldots,H_{n+1}(t);\zad_1(t),\ldots,\zad_n(t))}\!\!.
\end{eqnarray}
Eventually, one has to check that condition~{(\ref{small})} is
satisfied. Explicitly, one has to check that the solution obtained
$\{t\mapsto \ccad_n\lt,\, t\mapsto\zad_n\lt\}_{n\in\mathbb{N}}$ is
such that
\begin{eqnarray}
\extrat{\dot{\zad}_1(t)}\!\! \spa & \simeq & \spa 0,
\nonumber \\
& \vdots & \nonumber \\
\spa\spa\spa\!\! \extrat{\RRR_n^+ (\dot{\zad}_1(t),\ldots
,\dot{\zad}_{n-1}(t); \zad_1(t),\ldots, \zad_{n-1}(t))
-\dot{\zad}_n(t)}\!\! \spa & \simeq & \spa 0, \ \ \ n\ge 2,
\\
& \vdots & \nonumber
\end{eqnarray}

At this point, we have to show how the operator $\cad\lt$ can be
recovered, at each perturbative order, from the operator
$\ccad\lt$. To this aim, let us recall that
\begin{equation} \label{hermitian2}
\cad\lt = \ccad\lt-\sum_{k=1}^\infty
\frac{(-i)^k}{(k+1)!}\,\ad_{\cadint}^k\,\cad\lt .
\end{equation}
Now, notice that, if we substitute in
equation~{(\ref{hermitian2})} the power expansions
$\cad\lt=\sum_{n=1}^{\infty}\lambda^n\, \cad_n(t)$ and
$\ccad\lt=\sum_{n=1}^{\infty}\lambda^n\, \ccad_n(t)$, and we
single out the various perturbative orders, we conclude that the
$n$-th order, which on the l.h.s.\ is given simply by
$\lambda^n\,\cad_n(t)$, consists on the r.h.s.\ of
$\lambda^n\,\ccad_n(t)$ plus a function of
$\cad_1(t),\ldots,\cad_n(t)$ and $\int_0^t \cad_1(\ttt)\
\de\ttt,\ldots ,\int_0^t \cad_{n-1}(\ttt)\ \de\ttt$. Thus,
exploiting this fact, we can achieve an order by order solution.
Indeed, recalling definition~{(\ref{defRRR})}, one finds out that
$\cad\lt$ can be obtained order by order from $\ccad\lt$ by means
of the following recursive process:
\begin{eqnarray}
\cad_1(t) \spa & = & \spa \ccad_1(t),
\nonumber\\
& \vdots & \nonumber \\ \label{recpro*} \cad_n(t) \spa & = & \spa
\RRR_n^- \!\left(\cad_1(t),\ldots ,\cad_{n-1}(t); \int_0^t
\cad_1(\ttt)\ \de\ttt,\ldots ,\int_0^t \cad_{n-1}(\ttt)\ \de\ttt
\right)\nonumber \\ & + & \spa \ccad_n(t),\ \ \ \ n\ge 2,
\\
& \vdots & \nonumber
\end{eqnarray}
This recursive process allows to easily prove by induction that
the commutation relation $[\ccad\lt,\ho(t)]=0$ (or, equivalently,
$[\ccad\lt,P_m]=0$, $\forall m$) implies that
$[\cad\lt,\ho(t)]=0$. Thus equation~{(\ref{gen5})}, with
condition~{(\ref{costraint})}, is indeed equivalent to
equation~{(\ref{gen4})}, with condition~{(\ref{commute})}.\\ We
want to show now that there is also another recursive process
allowing to recover $\cad\lt$ from $\ccad\lt$ which less expensive
from the computational point of view.

To this aim, let us define the function $\avxp
:\mathbb{R}\rightarrow \mathbb{R}^+$ in the following way:
\begin{equation}
\avxp(x)=\frac{1}{x}\int_0^x e^t \,\de t
=\frac{e^x-1}{x}\ \ \; \mbox{for}\ x\neq 0,\ \ \
\avxp(0)=1.
\end{equation}
This function extends to an entire holomorphic function on the complex plane
which is given by
\begin{equation} \label{avxp}
\avxp(z)=\sum_{k=0}^{\infty}\frac{1}{(k+1)!}\, z^k.
\end{equation}
We recall that the function $\avxp$ is fundamental in the theory
of Lie groups since it is strictily related to the differential of
the exponential map, usually denoted by $\dexp$ (see
ref.~\cite{Raja}). In fact, given a Lie group $G$ with Lie algebra
$\mathfrak{g}$, identifying the tangent spaces at any point of $G$
and of $\mathfrak{g}$ with $\mathfrak{g}$ itself, one has:
\begin{equation}
\dexp(X)\,Y= \avxp(-X)\, Y,\ \ \ \ X,Y\in\mathfrak{g}.
\end{equation}
Besides, it is well known (see, for instance, ref.~\cite{Ahlfors})
that the the meromorphic function $1/\avxp$
admits the following expansion in the open disk of radius $2\pi$
centered at zero:
\begin{equation} \label{avxpinv}
\avxp(z)^{-1}=\sum_{k=0}^\infty\frac{\beta_k}{k!}\, z^k,
\end{equation}
where $\{\beta_0,\beta_1,\ldots\}$ are the {\it Bernoulli numbers},
namely the rational numbers defined recursively by
\begin{equation}
\beta_0=1,\  \  \left(\!\!\!\begin{array}{c} k+1 \\ 0 \end{array}
\!\!\!\right)\beta_0 + \cdots +\left(\!\!\! \begin{array}{c} k+1
\\ k \end{array} \!\!\! \right)\beta_k =0 \ \ \ k=1,2,\ldots\ .
\end{equation}
We recall also that
\begin{equation} \label{bernoulli}
\beta_{2k+1}=0, \ \ \ \frac{\beta_{2k}}{|\beta_{2k}|}=(-1)^{k+1},
\ \ \ k=1,2,\ldots\ .
\end{equation}
Now, notice that, according to definition~{(\ref{hermitian})} and
formula~{(\ref{avxp})}, we have:
\begin{equation}
\ccad\lt=\avxp\!\left(-i\,\ad_{\cadint}\right) \cad\lt.
\end{equation}
Then, by means of formula~{(\ref{avxpinv})}, we can write
\begin{eqnarray}
\cad\lt \spa & = & \spa
\avxp\!\left(-i\,\ad_{\cadint}\right)^{-1}\ccad\lt \nonumber\\
\label{invcc} & = & \spa
\sum_{k=0}^\infty\frac{(-i)^k\beta_k}{k!}\,\ad_{\cadint}^k\
\ccad\lt.
\end{eqnarray}
Let us observe that, as in the previous  case, if we substitute in
this equation the power expansions
$\cad\lt=\sum_{n=1}^{\infty}\lambda^n\, \cad_n(t)$ and
$\ccad\lt=\sum_{n=1}^{\infty}\lambda^n\, \ccad_n(t)$, and we
single out the various perturbative orders, we find that the
$n$-th order, which on the l.h.s.\ is given simply by
$\lambda^n\,\cad_n(t)$, consists on the r.h.s.\ of a function of
$\ccad_1(t),\ldots,\ccad_n(t)$ and $\int_0^t \cad_1(\ttt)\
\de\ttt,\ldots ,\int_0^t \cad_{n-1}(\ttt)\ \de\ttt$. Thus, again,
we can achieve an order by
order solution.\\
To this aim, given linear operators $X$ and
$Y_1,\ldots, Y_n$, $n\ge 1$, let us set
\begin{equation}
\BB_{n}^\pm(X; Y_1,\ldots ,Y_{n}) := \mp \sum_{m=1}^{n} \frac{(\pm
i)^m\beta_m}{m!} \!\!\!\!\! \sum_{\ \ \ k_1+\cdots +k_m=n}
\!\!\!\!\!\!\!\! \ad_{Y_{k_1}}\! \cdots\; \ad_{Y_{k_m}} X ;
\end{equation}
namely, we have that $\BB_1^\pm(X;Y_1)=\frac{i}{2}\,\ad_{Y_1} X$
(since $\beta_1=-1/2$) and, by virtue of
relations~({\ref{bernoulli})},
\begin{equation} \label{defBB}
\BB_{n}^\pm (X; Y_1,\ldots ,Y_{n}) = \frac{i}{2}\, \ad_{Y_n}\, X \pm
\sum_{m=1}^{\mathsf{p}(n)/2} \frac{|\beta_{2m}|}{2m!} \!\!\!\!\!\!\!
\sum_{\ \ \ k_1+\cdots +k_{2m}=n} \!\!\!\!\!\!\!\! \ad_{Y_{k_1}}\!
\cdots\; \ad_{Y_{k_{2m}}} X,
\end{equation}
for $n\ge 2$, where $\mathsf{p}(n)$ is equal to $n$ if $n$ is even
and to $n-1$ otherwise. Next, for $n\ge 2$, given linear operators
$X_1,\ldots, X_{n-1}$ and $Y_1,\ldots, Y_{n-1}$,  we can define
\begin{equation}
\BBB_n^\pm(X_1,\ldots ,X_{n-1}; Y_1,\ldots ,Y_{n-1})
:=\sum_{m=1}^{n-1} \BB_{n-m}^\pm(X_m;Y_1,\ldots ,Y_{n-m}),
\end{equation}
where the definition of  $\BBB_n^+(X_1,\ldots ,X_{n-1}; Y_1,\ldots
,Y_{n-1})$ will be used in the next section. Then, one finds out
that $\cad\lt$ can be obtained order by order from $\ccad\lt$ by
means of the following recursive process:
\begin{eqnarray}
\cad_1(t) \spa & = & \spa \ccad_1(t),
\nonumber\\
& \vdots & \nonumber \\ \label{recpro} \cad_n(t) \spa & = & \spa
\BBB_n^- \!\left(\ccad_1(t),\ldots ,\ccad_{n-1}(t); \int_0^t
\cad_1(\ttt)\ \de\ttt,\ldots ,\int_0^t \cad_{n-1}(\ttt)\ \de\ttt
\right)\nonumber \\ & + & \spa \ccad_n(t),\ \ \ \ n\ge 2,
\\
& \vdots & \nonumber
\end{eqnarray}
One can verify that this recursive process is computationally
convenient with respect to the recursive
process~{(\ref{recpro*})}. Specifically, it turns out that, from
the $4$-th perturbative order on, each step involves a smaller
number of terms in comparison with process~{(\ref{recpro*})}, with
a gain which increases step after step.


\section{The general case}
\label{generalcase}

In this section we will consider equation~{(\ref{eqcompl})} in
its full generality. This equation, taking into account definition~{(\ref{hermitian})},
can be rewritten as
\begin{equation} \label{equation}
\sum_{k=0}^\infty
\frac{i^k}{k!}\,\ad_{\zlt}^k\!\left(\rothamlt-\frac{1}{k+1}\,\dot{Z}\lt\right)
=  \cc\lt,
\end{equation}
where the operator $\cc\lt$ is linked to the operator $\clt$,
which appears in decomposition~{(\ref{gen2})} of the interaction
picture evolution operator, by the same relation that links the
operators $\ccad\lt$ and $\cad\lt$ introduced for studying the
adiabatic approximation; we rewrite it here for the sake of
clarity:
\begin{eqnarray}
\cc\lt \spa & := & \spa \avxp\!\left(-i\cltint\right)\clt
\nonumber \\ \label{hermitian*} & = & \spa \clt+\sum_{k=1}^\infty
\frac{(-i)^k}{(k+1)!}\,\ad_{\cltint}^k\,\clt .
\end{eqnarray}
We already know~(see section~{\ref{independent}}) that the
operator $\clt$ can be recovered from the operator $\cc\lt$ by
means of an order by order procedures. Thus, we can solve
equation~{(\ref{equation})} for $\cc\lt$ up to a given
perturbative order and obtain the perturbative expansion
of $\clt$ truncated at the same order parallely.\\
First of all, we
want to provide an interpretation of decomposition~{(\ref{gen2})}
which sheds light on its meaning. To this scope, notice that this
decomposition can be rewritten as
\begin{equation} \label{intpic}
\TC\lt=\TZ\lt^\dagger\, T\lt\,\TZ (\lambda;0),
\end{equation}
where
\begin{equation}
\TC\lt:=\exp\!\left(-i\cltint\right)\ \ \mbox{and}\ \
\TZ\lt:=\exp\!\left(-i\,\zlt\right).
\end{equation}
Now, formula~{(\ref{intpic})}
can be regarded as a passage to a further (generalized)
interaction picture performed on the Hamiltonian $\rothamlt$.
Indeed, assuming that $\TZ\lt$ satisfies the equation
\begin{equation}
i\,\dot{T}_Z\lt=\TZ\lt\,\zzz\lt,
\end{equation}
one finds that
\begin{equation}
i\,\dot{T}_C\lt=\left(\TZ\lt^\dagger\,\rotham\lt\,\TZ\lt-\zzz\lt\right)\TC\lt.
\end{equation}
Then, since
\begin{equation}
\zzz\lt= \sum_{k=0}^\infty
\frac{i^k}{(k+1)!}\,\ad_{\zlt}^k\,\dot{Z}\lt,
\end{equation}
equation~{(\ref{equation})} expresses precisely the fact that
$\cc\lt$ is the transformed Hamiltonian obtained by switching to
this new interaction picure; namely:
\begin{equation}
\cc\lt= \TZ\lt^\dagger\,\rotham\lt\,\TZ\lt-\zzz\lt.
\end{equation}
It follows that
\begin{equation}
\TC\lt=\exp\!\left(-i\sum_{n=1}^\infty\lambda^n\int_0^t
C_n(\ttt)\,\de\ttt\right)
\end{equation}
is nothing but the Magnus
expansion of evolution operator associated with the new interaction
picture Hamiltonian $\cc\lt$.

Let us now investigate the perturbative solutions of equation~{(\ref{equation})}.
Substituting the power expansions
\begin{equation} \label{t-power}
\rotham\lt = \sum_{n=1}^\infty \lambda^n\, \rotham_n(t),\ \
\cc\lt=\sum_{n=1}^{\infty}\lambda^n\, \cc_n(t),\ \  \zlt
=\sum_{n=1}^{\infty}\lambda^n\, Z_n(t),
\end{equation}
in equation~{(\ref{equation})}, we obtain an infinite set of
coupled equations that allows to compute the operators
$\{\cc_n(t)\}_{n\in\mathbb{N}}$, $\{Z_n(t)\}_{n\in\mathbb{N}}$. In
fact, in analogy with the time-independent case (compare with
definition~{(\ref{ef})}), given linear operators $X$ and $Y$, for
$n\ge 1$, let us set
\begin{equation}
\hat{\mathcal{G}}_{n}(X,Y; Z_1,\ldots ,Z_{n}) := \sum_{m=1}^{n}
\frac{i^m}{m!} \!\!\!\!\!\!\! \sum_{\ \ \ k_1+\cdots +k_m=n}
\!\!\!\!\!\!\!\!\! \ad_{Z_{k_1}} \cdots\
\ad_{Z_{k_m}}\!\!\left(X-\frac{Y}{m+1}\right)\!.
\end{equation}
Then we can define $\hG_n(\rotham_1(t),\ldots ,\rotham_n(t) ;
Z_1(t),\ldots ,Z_{n-1}(t); \dot{Z}_1(t),\ldots ,\dot{Z}_{n-1}(t))$
as
\begin{equation}
\sum_{m=1}^{n-1} \hat{\mathcal{G}}_{n-m}(\rotham_m(t),\dot{Z}_m(t);Z_1(t),\ldots
,Z_{n-m}(t))+\rotham_n(t), \ \ \ n\ge 2.
\end{equation}
With these notations, one can write the sequence of coupled
equations which gives a perturbative solution of equation~{(\ref{equation})}
as follows:
\begin{eqnarray}
\dot{Z}_1(t) \spa & = & \spa \rotham_1(t)-\cc_1\lt,
\nonumber\\
& \vdots & \nonumber \\
\dot{Z}_n(t) \spa & = & \spa \hG_n\! \left(\rotham_1(t),\ldots
,\rotham_n(t); Z_1(t),\ldots, Z_{n-1}(t); \dot{Z}_1(t),\ldots,
\dot{Z}_{n-1}(t)\right)
\nonumber\\ \label{eqset}
& - & \spa \cc_n(t),\ \ \ \ \ n\ge 2,
\\
& \vdots & \nonumber
\end{eqnarray}
It is clear that, as in the time-independent case, this infinite
set of equations can be solved recursively. Moreover, recalling
the recursive process~{(\ref{recpro})} that allows to obtain, at
each perturbative order, the expression of $\clt$, one can
calculate order by order both the operators
$\{C_n(t)\}_{n\in\mathbb{N}}$ and $\{Z_n(t)\}_{n\in\mathbb{N}}$.
Indeed, integrating with respect to time each equation in the
sequence~{(\ref{eqset})} and combining the new sequence of
equations so obtained with the recursive process~{(\ref{recpro})}
(or~{(\ref{recpro*})}), we find:
\begin{eqnarray}
\spa\spa\spa\spa Z_1(t) \spa & = &  \spa \int_0^t\!
\left(\rotham_1(\ttt)-\cc_1(\ttt)\right)\de \ttt +  Z_1  ,
\nonumber\\
\spa\spa\spa\spa C_1(t) \spa & = & \spa \cc_1(t),
\nonumber\\
& \vdots &
\nonumber\\
\spa\spa\spa\spa Z_n(t) \spa & = & \spa \int_0^t \!\left(\hG_n
\big(\ldots ,\rotham_n(\ttt);\ldots, Z_{n-1}(\ttt); \ldots,
\dot{Z}_{n-1}(\ttt)\big)-\cc_n(\ttt)\right)\de\ttt + Z_n ,
\nonumber\\
\spa\spa\spa\spa C_n(t) \spa & = & \spa \BBB_n^-\!\left(\ldots
,\cc_{n-1}(t); \ldots ,\int_0^t C_{n-1}(\ttt)\ \de\ttt \right)\! +
\cc_n(t) ,\ \ \ \ n\ge 2,
\\
& \vdots & \nonumber
\end{eqnarray}
This time, differently from the equations obtained in
section~{\ref{independent}} for the time-independent case, at each
perturbative order we have a {\it couple} of equations. The
solution of the first couple of equations is obtained by simply
choosing the arbitrary operator-valued function $t\mapsto\cc_1(t)$
and the arbitrary operator $Z_1$; similarly, the solution of the
$n$-th couple of equations, for $n\ge 2$, involves the previously
computed functions $t\mapsto Z_1(t),\ldots, t\mapsto Z_{n-1}(t)$
and requires only the choice of the arbitrary operator-valued
function $t\mapsto \cc_n(t)$ and of the arbitrary operator $Z_n$.
This choice can be fitted according to computational convenience.
Notice that, in particular, the choice of the operators
$\{Z_n\}_{n\in\mathbb{N}}$ determines the initial condition
\begin{equation}
\zl\equiv Z(\lambda ;0)=\sum_{n=1}^\infty \lambda^n\, Z_n.
\end{equation}
We will see soon that one can set quite natural conditions which
fix a unique choice of these arbitrary quantities uniquely. Before
doing this, we want to show that it is possible to write a
sequence of equations which is equivalent and structurally similar
to the one given above but such that the solution of the $n$-th
couple of equations
--- the one determining $C_n(t)$ and $Z_n(t)$ ---
does not involve explicitly the operators $\dot{Z}_1(t),\ldots ,
\dot{Z}_{n-1}(t)$.

The first step is to rewrite equation~{(\ref{equation})} as
\begin{equation}
\avxp\!\left(i\,\ad_{\zlt}\right)\dot{Z}\lt=
\exp\!\left(i\,\ad_{\zlt}\right)\rotham\lt-\cl.
\end{equation}
Next, using the relation
\begin{equation}
\avxp(z)^{-1}\exp(z)=\avxp(-z)^{-1},
\end{equation}
we have:
\begin{eqnarray}
\dot{Z}\lt \spa & = & \spa
\avxp\!\left(i\,\ad_{\zlt}\right)^{-1}\left(
\exp\!\left(i\,\ad_{\zlt}\right)\rotham\lt-\cl\right) \nonumber\\
\label{Equation} & = & \spa \sum_{k=0}^\infty
\frac{i^k\beta_k}{k!}\,\ad_{\zlt}^k\left((-1)^k\rotham\lt-\cl\right).
\end{eqnarray}
Recalling the fact that $\beta_0=1,\ \beta_1=-1/2$
and relations~{(\ref{bernoulli})},
we can further simplify equation~{(\ref{Equation})}:
\begin{eqnarray}
\dot{Z}\lt
\spa & = & \spa \rotham\lt-\cl +
\frac{i}{2}\,\ad_{\zlt}\!\left(\rotham\lt+\cl\right) \nonumber\\
& + & \spa
\sum_{k=1}^\infty (-1)^k
\frac{\beta_{2k}}{2k!}\,\ad_{\zlt}^{2k}\left(\rotham\lt-\cl\right)
=
\frac{i}{2}\,\ad_{\zlt}\!\left(\rotham\lt+\cl\right)
\nonumber \\
& + & \spa
\left(1-\sum_{n=1}^\infty
\frac{|\beta_{2k}|}{2k!}\,\ad_{\zlt}^{2k}\right)\!
\left(\rotham\lt-\cl\right).
\label{Equation*}
\end{eqnarray}

Notice that in the time-independent case ($\hilt\equiv\hil$),
assuming as in section~{\ref{independent}} that $[\cl,\ho]=0$ and
$\zlt=\exp(i\ho t)\,\zl\,\exp(-i\ho t)$, hence
\begin{equation}
\dot{Z}\lt=-i\,\Ad_{\exp(i\ho t)}\,\ad_{\zl}\,\ho,
\end{equation}
from equation~{(\ref{Equation*})}, using relation~{(\ref{relad})},
we find:
\begin{eqnarray}
-i\,\ad_{\zl}\,\ho \spa & = & \spa \hil - \cl +
\frac{i}{2}\,\ad_{\zlt}\!\left(\hil+\cl\right) \nonumber\\
\label{Equation**} & - & \spa \sum_{k=1}^\infty
\frac{|\beta_{2k}|}{2k!}\,\ad_{\zlt}^{2k}\left(\hil-\cl\right).
\end{eqnarray}
This equation yields another order by order solution procedure
with respect to the one described in section~{\ref{independent}}.
Anyway, we will not insist on this point and we leave the details
to the reader.

Returning to the general time-dependent case, from
equation~{(\ref{Equation*})}, we can obtain an infinite set of
equations which allows to give perturbative expressions of $\zlt$,
$\cl$ and can be solved recursively. To this aim, given linear
operators $X$, $Y$ and $Z_1,\ldots,Z_n$, let us set
\begin{equation}
\spa\BB_{n}(X,Y; Z_1,\ldots ,Z_{n}) := \sum_{m=1}^{n}
\frac{i^m\beta_m}{m!} \!\!\!\!\!\!\!\! \sum_{\ \ \ k_1+\cdots
+k_m=n} \!\!\!\!\!\!\!\!\!\!\! \ad_{Z_{k_1}}\! \cdots\;
\ad_{Z_{k_m}}\!\!\left((-1)^m X - Y\right)\! ,
\end{equation}
for $n\ge 1$; namely,
$\BB_1(X,Y;Z_1)=\frac{i}{2}\,\ad_{Z_1}\!\left(X+Y\right)$ and,
recalling definition~{(\ref{defBB})},
\begin{eqnarray}
\BB_{n}(X,Y; Z_1,\ldots ,Z_{n}) \spa & := & \spa
\frac{i}{2}\,\ad_{Z_n}\!\left(X+Y\right) \nonumber \\
& + & \spa\! \sum_{m=1}^{\mathsf{p}(n)/2} \frac{|\beta_{2m}|}{2m!}
\!\!\!\!\!\!\! \sum_{\ \ \ k_1+\cdots +k_{2m}=n}
\!\!\!\!\!\!\!\!\! \ad_{Z_{k_1}}\! \cdots\;
\ad_{Z_{k_{2m}}}\!\left(Y - X\right)
\nonumber \\
& = & \spa \BB_{n}^-(X; Z_1,\ldots ,Z_{n}) + \BB_{n}^+(Y;
Z_1,\ldots ,Z_{n}),
\end{eqnarray}
for $n\ge 2$, where $\mathsf{p}(n)$ is equal to $n$ if $n$ is even
and to $n-1$ otherwise. Then we can define, for $n\ \ge 2$,
\[
\BBB_n\equiv\BBB_n(\rotham_1(t),\ldots ,\rotham_{n-1}(t) ;
\cc_1(t),\ldots ,\cc_{n-1}(t); Z_1(t),\ldots ,Z_{n-1}(t))
\]
as
\begin{eqnarray}
\spa\spa\BBB_n \spa & := & \spa \sum_{m=1}^{n-1}
\BB_{n-m}(\rotham_m(t),\cc_m(t);Z_1(t),\ldots ,Z_{n-m}(t)) \nonumber \\
\spa & = & \spa \BBB_n^-(\ldots ,\rotham_{n-1}(t) ; \ldots
,Z_{n-1}(t))   +   \BBB_n^+ (\ldots ,\cc_{n-1}(t); \ldots
,Z_{n-1}(t)).
\end{eqnarray}
At this point, inserting the power expansions~{(\ref{t-power})} in
equation~{(\ref{Equation})} and equating the terms of the same
order in the perturbative parameter $\lambda$, we obtain the
following sequence of equations:
\begin{eqnarray}
\dot{Z}_1(t) \spa & = & \spa \rotham_1(t)-\cc_1\lt,
\nonumber\\
C_1(t)\spa & = & \spa \cc_1(t), \nonumber\\
& \vdots & \nonumber \\
\dot{Z}_n(t) \spa & = & \spa \BBB_n(\rotham_1(t),\ldots
,\rotham_{n-1}(t);\cc_1(t),\ldots,\cc_{n-1}(t); Z_1(t),\ldots ,
Z_{n-1}(t)) \nonumber \\
& + & \spa \rotham_n(t)- \cc_n(t), \nonumber \\
\label{system} C_n(t) \spa & = & \spa \BBB_n^-
\!\left(\cc_1(t),\ldots ,\cc_{n-1}(t); \int_0^t C_1(\ttt)\
\de\ttt,\ldots ,\int_0^t C_{n-1}(\ttt)\ \de\ttt \right)\nonumber
\\ & + & \spa \cc_n(t),\ \ \ \ n\ge 2,
\\
& \vdots & \nonumber
\end{eqnarray}
Again, this sequence of equations can be solved recursively and
the solution of the $n$-th couple of equations requires simply the
choice of an arbitrary operator-valued function
($t\mapsto\cc_n(t)$) and of an arbitrary operator ($Z_n$).\\
An important class of solutions is determined by the condition
\begin{equation}
\cc_1(t)=\cc_1(0)\equiv \cc_1,\ldots,
\cc_n(t)=\cc_n(0)\equiv\cc_n,\ldots\ \ \forall t.
\end{equation}
This condition is equivalent to the following:
\begin{equation} \label{conc}
C_1(t)=C_1(0)\equiv  C_1,\ldots,  C_n(t)=C_n(0)\equiv C_n,\ldots\
\ \forall t.
\end{equation}
Moreover, if this condition holds, we have:
\begin{equation}
C_1=\cc_1,\ldots,  C_n=\cc_n,\ldots\ .
\end{equation}
Then the solution of the first equation, namely
\begin{equation}
Z_1(\{C_1,Z_1\};t)=\int_0^t \rotham_1(\ttt)\,\de\ttt -t\, C_1+Z_1,
\end{equation}
is fixed by the choice of the `arbitrary constants' $C_1$ and
$Z_1$. Inductively, the solution of the $n$-th equation is
obtained by substituting the previously chosen arbitrary constants
$C_1,\ldots,C_{n-1}$ and the solutions
\[
t\mapsto Z_1(\{C_1,Z_1\},t),\ldots,t\mapsto
Z_{n-1}(\{C_k,Z_k\}_{k=1}^{n-1};t)
\]
of the first $n-1$ equations --- that are fixed by the choice of
the additional arbitrary constants $Z_1,\ldots,Z_{n-1}$ --- in the
formula
\begin{eqnarray}
\spa Z_n(\{C_k,Z_k\}_{k=1}^n;t) \spa & = &  \spa \int_0^t\!
\BBB_n\!\left(\ldots ,\rotham_{n-1}(\ttt);\ldots,C_{n-1}; \ldots ,
Z_{n-1}(\{C_k,Z_k\}_{k=1}^{n-1};\ttt)\right)\! \de\ttt \nonumber\\
\label{zeta} & + & \spa \int_0^t\rotham_n(\ttt)\,\de\ttt -t\,
C_n+Z_n,
\end{eqnarray}
which involves the $n$-th order arbitrary constants $C_n$ and
$Z_n$.

Now, as anticipated, we will give suitable conditions that fix the
arbitrary constants $\{C_n,Z_n\}_{n\in\mathbb{N}}$ --- hence, a
solution of the system~{(\ref{system})} --- uniquely up to a
certain perturbative order $\enne\in\mathbb{N}$.\\
To this aim, it will be convenient to introduce the following
notations. First, for the sake of brevity, let us define
\begin{equation}
\JJJ_n(\ldots
):=
\BBB_n(\ldots,\rotham_{n-1}(t);\ldots,C_{n-1};\ldots,Z_{n-1}(t))+
\rotham_n(t), \ \ \ n\ge 2.
\end{equation}
Next, given an analytic function $\lambda\mapsto
f(\lambda)=\sum_{n=0}^{\infty}\lambda^n f_n$, we will set
\begin{equation}
f\enneind (\lambda):= \sum_{n=0}^{\enne} \lambda^n f_n.
\end{equation}
Moreover, given another analytic function $\lambda\mapsto
h(\lambda)$, we will set:
\begin{equation}
f(\lambda)\app h(\lambda)\ \ \
\stackrel{\mathrm{def}}{\Longleftrightarrow}\ \ \
f\enneind(\lambda)=h\enneind(\lambda).
\end{equation}
Finally, given $\bar{t}\in(0,\infty]$, we will say that a function
$t\mapsto F(t)$ has {\it zero average over the time span}
$[0,\bar{t}\,]$ if
\begin{equation} \label{average}
\langle F(\cdot)\rangle_{\bar{t}}:=
\lim_{t\rightarrow\bar{t}}\,\frac{1}{t}\int_0^t F(\ttt)\,\de
\ttt=0,
\end{equation}
where, obviously, the limit is essential in the previous
definition only in the case where $\bar{t}=\infty$.
For $\bar{t}<\infty$, equation~{(\ref{average})} expresses the fact that
the function $t\mapsto F(t)$ has a `purely oscillatory behavior' in the interval $[0,\bar{t}\,]$.\\
Then, fixed a certain perturbative order $\enne$ and $\tau\in \left]0\,
,\infty\right[\,$, we set the following conditions:
\begin{description}
\item[C1]\ the following relation holds:
\begin{equation} \label{cond1}
Z_n (\tau)=Z_n, \ \ \ n=1,\ldots,\enne ;
\end{equation}
\item[C2]\ the operator-valued function $t\mapsto Z_n\lt,\ n=1,\ldots,\enne$,
has zero average over the time span $[0,\tau]$.
\end{description}
Condition~{\bf C1} is equivalent to the following:
\begin{equation}
Z\enneind(\lambda ;\tau)=Z\enneind(\lambda).
\end{equation}
Thus, since
\begin{equation}
\rotev (\lambda ;\tau) \app \exp\!\left(-i\,Z\enneind(\lambda
;\tau)\right)\,
\exp\!\left(-i\,C\enneind(\lambda)\,\tau\right)\,\exp\!
\left(iZ\enneind(\lambda)\right),
\end{equation}
condition~{\bf C1} implies that
\begin{eqnarray}
\rotev (\lambda ;\tau) \spa & \app & \spa
\exp\!\left(-i\,Z\enneind(\lambda)\right)\,
\exp\!\left(-i\,C\enneind(\lambda)\,\tau\right)\,\exp\!
\left(i\,Z\enneind(\lambda)\right),
\nonumber \\
& \app & \spa
\exp\!\left(-i\,\zl\right)\,\exp\left(-i\,\cl\,\tau\right)\,
\exp\!\left(i\,\zl\right).
\end{eqnarray}
If we are able to show that condition~{\bf C1} can indeed be
satisfied, this result has the following interpretation. There
exists a hermitian operator $\mathfrak{H}(\lambda)$, depending
analytically on the perturbative parameter $\lambda$, such that
the 1-parameter group of unitary operators generated by it
interpolates, up to the $\enne$-th perturbative order, the
evolutor $\rotev\lt$ at $t=\tau$; namely:
\begin{equation}
\rotev (\lambda ;\tau)\app
\exp\!\left(-i\,\mathfrak{H}(\lambda)\,\tau\right).
\end{equation}
Indeed, this relation is satisfied if we set
\begin{equation}
\mathfrak{H}(\lambda)= \exp\!\left(-i\,\zl\right) \cl\,
\exp\!\left(i\,\zl\right).
\end{equation}
Now, observe that, denoted by $\{\tauc_n,\,\tauz_n\}_{n=1}^\enne$ an item
of the first $\enne$ arbitrary operator constants satisfying
conditions~{\bf C1} and {\bf C2}, applying condition {\bf C1} to
formula~{(\ref{zeta})} yields
$\tauc_1=\langle\rotham(\cdot)\rangle_\tau$ and
\begin{equation} \label{eqc}
\tauc_n = \frac{1}{\tau} \int_0^\tau\! \JJJ_n\!\left(\ldots
,\rotham_n(t);\ldots,\tauc_{n-1}; \ldots ,
Z_{n-1}(\{\tauc_k,\tauz_k\}_{k=1}^{n-1};t)\right) \de t,
\end{equation}
for $n=2,\ldots ,\enne$. Thus condition~{\bf C1} determines the
operators $\{\tauc_n\}_{n=1}^{\enne}$ uniquely for a fixed
$\enne$-tuple $\{\tauz_n\}_{n=1}^\enne$. Suppose that
$t\mapsto\rotham\lt$ is periodic up to the $\enne$-th perturbative
order, with period $\period$
--- $\rotham\enneind(\lambda;t)=\rotham\enneind(\lambda;t+\period),\ \forall t$
--- or equivalently
\begin{equation}
\rotham_n(t)=\rotham_n(t+\period),\ \ \ \forall t,\ \forall
n\in\{1,\ldots,\enne\}.
\end{equation}
Then the functions $t\mapsto Z_1(C_1,Z_1;t),\ldots,t\mapsto
Z_\enne(\{C_k,Z_k\}_{k=1}^{\enne-1};t)$ are periodic, with period
$\period$, if and only if
$C_1=\tauc_1,\ldots,C_\enne=\tauc_\enne$, with $\tau=m_0\period$,
for some $m_0\in\mathbb{N}$. In fact, we have:
\begin{eqnarray}
\spa Z_1(C_1,Z_1;t+\period) \spa & = & \spa \int_0^t
\rotham_1(\ttt)\,\de\ttt+\int_t^{t+\period}\!\!\spa
\rotham_1(\ttt)\,\de\ttt-(t+\period)\,C_1+Z_1 \nonumber\\
\spa\spa\spa\spa\spa *\,\rotham_1(\cdot)\ \mbox{periodic}\,* & = &
\spa\int_0^t
\rotham_1(\ttt)\,\de\ttt-t\,C_1+Z_1+\left(\int_0^{\period}
\rotham_1(\ttt)\,\de\ttt -\period\,C_1\right).
\end{eqnarray}
Hence, the function $t\mapsto Z_1(C_1,Z_1;t)$ is periodic if and
only if
\begin{equation}
C_1=\langle \rotham_1(\cdot)\rangle_\period=\tauc_1,\ \ \
\mbox{for}\ \tau=m_0\period,\ m_0\in\mathbb{N}.
\end{equation}
Next, reasoning by induction and using the fact that the function
\[
t\mapsto \JJJ_n\!\left(\rotham_1(t),\ldots
,\rotham_n(t);\ldots,C_{n-1}; Z_1(C_1,Z_1;t),\ldots ,
Z_{n-1}(\{C_k,Z_k\}_{k=1}^{n-1};t)\right)
\]
is periodic, with period $\period$, if the functions
\[
\rotham_1(\cdot),\ldots ,\rotham_n(\cdot);\
Z_1(C_1,Z_1;\cdot),\ldots , Z_{n-1}(\{C_k,Z_k\}_{k=1}^{n-1};\cdot)
\]
are periodic, with period $\period$, one proves the claim.\\
It follows that, if $t\mapsto\rotham\lt$ is periodic up to the
$\enne$-th order, then, setting $\tau=m_0\period$ in
condition~{\bf C1} for some nonzero positive integer $m_0$, we
have:
\begin{equation}
\rotev(\lambda;m\period)\app\exp(-i\,\mathfrak{H}(\lambda)\,m\period),\
\ \ m=1,2,\ldots\ .
\end{equation}

Condition~{\bf C2} is equivalent to the condition that the
operator-valued function $t\mapsto Z\enneind(\lambda;t)$ has zero
average over the time span $[0,\tau]$. Besides, applying
condition~{\bf C2}, namely
\begin{equation}
\frac{1}{\tau}\int_0^\tau Z_n(t)\, \de t=0,\ \ \ n=1,\ldots,\enne,
\end{equation}
to formula~{(\ref{zeta})} yields
$\tauz_1=-\langle\int_0^{(\cdot)}\rotham_1(\ttt)\,\de\ttt\,\rangle_\tau
+\frac{1}{2}\,\tau\,\tauc_1$ and
\begin{eqnarray}
\spa \tauz_n \spa & = & \spa -\frac{1}{\tau} \int_0^\tau\!
\!\!\left(\int_0^t\! \JJJ_n\!\left(\ldots
,\rotham_n(\ttt);\ldots,\tauc_{n-1}; \ldots ,
Z_{n-1}(\{\tauc_k, \tauz_k\}_{k=1}^{n-1};\ttt)\right)
\de\ttt\right) \de t \nonumber\\ \label{eqz} & + & \spa
\frac{1}{2}\,\tau\, \tauc_n , \ \ \ \ n=2,\ldots,\enne.
\end{eqnarray}
Hence, as claimed before, there is a unique $\enne$-tuple
$\{\tauc_n,\tauz_n\}_{n=1}^{\enne}$, determined by formulae~{(\ref{eqc})}
and~{(\ref{eqz})}, satisfying conditions {\bf C1} and {\bf C2}.\\

Now, let us suppose that the following limits exist:
\begin{eqnarray}
\infc_1\spa & := & \spa \lim_{\tau\rightarrow\infty}\tauc_1=
\lim_{\tau\rightarrow\infty}\frac{1}{\tau}
\int_0^\tau \rotham_1(t)\ \de t ,
\nonumber \\
\infz_1 \spa & := & \spa \lim_{\tau\rightarrow\infty}
\left(-\frac{1}{\tau}\int_0^\tau \!\left(
\int_0^t\rotham_1(\ttt)\ \de\ttt\right)\de t
+\frac{1}{2}\,\tau\,\infc_1 \right),
\nonumber \\
& \vdots &
\nonumber \\
\infc_\enne  \spa
& := &
\spa \lim_{\tau\rightarrow\infty} \left(
\frac{1}{\tau} \int_0^\tau\!
\JJJ_\enne\!\left(\ldots
;\ldots,\!\infc_{\enne -1}; \ldots ,
Z_{\enne-1}(\{\infc_k, \infz_k\}_{k=1}^{\enne -1}; t)\right)
\de t \right),
\nonumber \\
\infz_\enne \spa & := & \spa \lim_{\tau\rightarrow\infty} \bigg(\!
-\frac{1}{\tau} \int_0^\tau\! \!\!\left(\int_0^t\!
\JJJ_\enne\!\left(\ldots ;\ldots; \ldots ,
Z_{\enne -1}(\{\infc_k, \infz_k\}_{k=1}^{\enne -1};\ttt)\right)
\de\ttt\right) \de t \nonumber\\ & + & \spa
\frac{1}{2}\,\tau\,\infc_\enne\bigg).
\label{limits}
\end{eqnarray}
Then, the following relations hold:
\begin{equation} \label{relz}
\lim_{t\rightarrow\infty}
\frac{1}{t}\,Z_n(\{C_k,Z_k\}_{k=1}^n;t)=0,  \ \ \ \
n=1,\ldots,\enne,
\end{equation}
\begin{equation} \label{relzz}
\lim_{t\rightarrow\infty}
\frac{1}{t}\int_0^t Z_n(\{C_k,Z_k\}_{k=1}^{n};\ttt)\ \de \ttt =0, \ \ \ \
n=1,\ldots ,\enne,
\end{equation}
where $C_1=\infc_1,\, Z_1=\infz_1,
\ldots,C_\enne=\infc_\enne,\,Z_\enne=\infz_\enne$.
Conversely, if relations~{(\ref{relz})} and~{(\ref{relzz})}
hold for some operators
$\{C_k,Z_k\}_{k=1}^\enne$, then the limits~{(\ref{limits})} exist
and $C_1=\infc_1,\,Z_1=\infz_1,\ldots,C_\enne=\infc_\enne,
\,Z_\enne=\infz_\enne$.
Thus we can include the set $\{\infc_k,\infz_k\}_{k=1}^\enne$ among the
sets of operator constants determined by conditions~{\bf C1} and~{\bf C2} if
we allow $\tau=\infty$ and we rewrite condition~{\bf C1}:
\begin{description}
\item[C1]\ the following relation holds:
\begin{equation} \label{cond1*}
\lim_{t\rightarrow\tau}\frac{1}{t}\,Z_n(t)=
\lim_{t\rightarrow\tau}\frac{1}{t}\, Z_n, \ \ \  n=1,\ldots,\enne,
\end{equation}
\end{description}
while condition~{\bf C2} remains unchanged. Indeed,
relation~{(\ref{cond1*})} reduces to equation~{(\ref{cond1})} if
$\tau<\infty$ and to relation~{(\ref{relz})} if $\tau=\infty$.
Moreover, relation~{(\ref{relzz})} expresses the fact that the
function $t\mapsto Z_n(\lambda;t),\ n=1,\ldots,\enne$, has zero
average over the time span $[0,\infty]$.

At this point, considering the time-independent case ---
$\ho(t)\equiv\ho$ and $\hilt\equiv\hi(\lambda)$ ---
it is natural to ask what is the relation between the solution
associated with the arbitrary constants
$\{\infc_n,\infz_n\}_{n\in\mathbb{N}}$ (if they exist) discussed
in this section and the solutions obtained in
section~{\ref{independent}}. As we have done in that section,
we will assume that the unperturbed Hamiltonian $\ho$ has a pure point spectrum
$E_1,E_2,\ldots$ and we will denote by $P_1,P_2,\ldots$ the associated eigenprojectors.
Then, we want to prove that:
\begin{enumerate}
\item
for any $\enne\in\mathbb{N}$, the limits~{(\ref{limits})} --- thus
the set of operator constants $\{\infc_n,\,\infz_n\}_{n\in\mathbb{N}}$ ---
exist;

\item the minimal solution $\{\minc_n,\,\minz_n\}_{n\in\mathbb{N}}$
of the sequence of equations~{(\ref{general})}, i.e.\ the solution obtained imposing
condition~{(\ref{minsol})}, satisfies the relation
\begin{equation}
\minc_n=\infc_n, \ \ \ \minz_n=\infz_n,\ \ \forall n\in\mathbb{N};
\end{equation}

\item the operator-valued function $t\mapsto \zlt=\sum_{n=1}^\infty \lambda^n\,Z_n(t)$
which verfies conditions {\bf C1} and {\bf C2}, with $\tau=\infty$, is
such that
\begin{equation}
Z_n(t)\equiv Z_n(\{\infc_k,\infz_k\}_{k=1}^{n-1};t)=
e^{i\ho t}\,\infz_n\,e^{-i\ho t},\ \ \ \forall n\in\mathbb{N}.
\end{equation}
\end{enumerate}
In fact, we know that $\{\minc_n(t)=\minc_n,\,\minz_n(t)= e^{-i\ho
t}\,\,\minz_n\,e^{i\ho t}\}_{n\in\mathbb{N}}$ is a solution of the
sequence of equations~{(\ref{system})} with
condition~{(\ref{conc})}. Observe that
\begin{equation}
\lim_{t\rightarrow\infty}\frac{1}{t}\,\minz_n(t)=0,\ \ \ \forall n\in\mathbb{N},
\end{equation}
thus condition~{\bf C1}, with $\tau=\infty$ and $\enne$ arbitrary, is satisfied.
Moreover, given a linear
operator $X$, we have:
\begin{eqnarray}
X(t) \spa & := & \spa e^{i\ho t}\, X\, e^{-i\ho t}
\nonumber \\
*\,\mbox{since}\ [\infra{X},\ho]=0\,*  & = & \spa \infra{X} +
e^{i\ho t} \extra{X} e^{-i\ho t}
\nonumber \\
& = & \spa \infra{X} +  \sum_{j\neq l} e^{i(E_j-E_l)t}\, P_j\, X
\, P_l;
\end{eqnarray}
hence:
\begin{equation}
\langle X(\cdot)\rangle_\infty=\infra{X}.
\end{equation}
By condition~{(\ref{minsol})}, it follows that $\langle
\minz_n(\cdot)\rangle_\infty=0$, $n=1,2,\ldots$, thus condition~{\bf C2},
with $\tau=\infty$ and $\enne$ arbitrary, is
satisfied. Then, by the uniqueness of the solution satisfying
conditions~{\bf C1} and {\bf C2}, for a given
$\tau\in ]0,\infty]$ and up to a certain perturbative order $\enne$, we must conclude that
\begin{equation}
\minc_n=\infc_n \ \ \mbox{and}\ \
\minz_n(t)=Z_n(\{\infc_k,\infz_k\}_{k=1}^{n-1};t) ,\ \ \ \forall
n\in\mathbb{N}.
\end{equation}
Consequently, for any $n\in\mathbb{N}$, we have:
\begin{equation}
\minz_n=\minz_n(0)=Z_n(\{\infc_k,\infz_k\}_{k=1}^{n-1};0)=\infz_n
\end{equation}
and
\begin{eqnarray}
Z_n(\{\infc_k,\infz_k\}_{k=1}^{n-1};t)\spa & = & \spa
\minz_n(t)\nonumber\\
& = & \spa e^{-i\ho t}\,\,\minz_n\,e^{i\ho t}\nonumber \\ & = &
\spa e^{-i\ho t}\,\,\infz_n\,e^{i\ho t}.
\end{eqnarray}
This completes our proof.


\end{document}